\documentclass[DIV=calc, paper=a4, fontsize=11pt, onecolumn]{scrartcl}
\usepackage{amsfonts,amssymb,amsmath}
 \usepackage{lineno,hyperref}
 \usepackage{textcomp,mathcomp}
 \usepackage{caption,subcaption}
\usepackage{graphicx}
\usepackage{float}
\usepackage{tabularx}
\usepackage{authblk} % for author affiliations
\usepackage{arydshln,leftidx,mathtools}
\usepackage{accents}
\usepackage{xcolor}
\usepackage[noend]{algpseudocode}
\modulolinenumbers[1]
\usepackage{xcolor}   

\usepackage[english]{babel}
\usepackage[utf8]{inputenc}
\usepackage{parskip}
\usepackage{multirow}
\usepackage[top=2.5cm, left=3cm, right=3cm, bottom=4.0cm]{geometry}

\title{Host-feeding preferences and temperature shape the dynamics of West Nile virus: a mathematical model of assessing the abatement planning }
\author[1]{Suman Bhowmick\footnote{Corresponding Author}}
\author[4]{Megan Fritz}
\author[1, 2, 3]{Rebecca Lee Smith}
\affil[1]{Department of Pathobiology, University of Illinois at Urbana-Champaign, Urbana, Illinois, USA}
\affil[2]{Carl R. Woese Institute for Genomic Biology, University of Illinois at Urbana-Champaign, Urbana, Illinois, USA}
\affil[3]{Carle Illinois College of Medicine, University of Illinois at Urbana-Champaign, Urbana, Illinois, USA}
\affil[4]{Department of Entomology, Institute for Advanced Computer Studies, University of Maryland, USA}

\begin{document}
\maketitle
\tableofcontents
\section{Highlight}
\begin{itemize}
\item Mosquito feeding index is an important parameter that practically dictates the course of the WNV dynamics
\item Continuous introduction of infected agents into the system can keep the epidemic of WNV alive
\item Simply spraying adulticide is not enough to eliminate  cases of WNV given the nonlinear functional relationship between the feeding index and the efficacy rate of adulticide
\end{itemize}

%%%%%%%%%%%%%%%%%%%%%%%%%%%%%%%
\section{Abstract}
West Nile virus (WNV) is prevalent in the United States but it shows considerable divergence in transmission patterns and spatio-temporal intensity. 
It is to be noted that the mechanism that drives the transmission potential of WNV is described by the abilities of host species to maintain and disseminate the pathogens pertinent with different eco-epidemiological factors that have an influence on the contact rates amongst the interacting species.  
There is growing evidence that several vectors exhibit strong feeding preferences towards different host communities. 
In our research study, we construct a process based weather driven ordinary differential equation (ODE) model to understand the impact of one vector species (\textit{Culex pipiens}), preferred avian and non-preferred human hosts and compared it surveillance data for the \textit{Culex pipiens} complex collected in Cook County, Illinois, USA.
In our mechanistic model, we also demonstrate that adulticide treatments produced significant reductions in the \textit{Culex pipiens} population.
We take into account the feeding index that can be described as the ratio between observed frequency of mosquitoes feeding on one host compared to another host, divided by the expected frequency of mosquitoes feeding on these two hosts based on the presence of the particular hosts to develop this transmission model for WNV.
We also include continuous introduction of infected agents into the model during the simulations as the introduction of WNV is not a single event phenomenon. 
                 Sensitivity analysis demonstrates that feeding index and rate of introduction of infected agents are two important factors beside the efficacy of adulticide. 
We derive an analytic form of $R_0$ to predict the conditions under which there will be an outbreak of WNV and the relationship between the feeding index and the efficacy of adulticide is highly nonlinear.
Our findings demonstrate that the interplay between the feeding index and mosquito abatement strategy is rather a complex phenomenon and it induces a heterogeneous contact rates that should be included while modelling multi-host, multi-vector transmission model.

\section{Introduction}
%\textcolor{red}{What is WNV and its spread in the States}\\
%West Nile virus (WNV) is a viral infection that is transmitted to humans and mammals through the bites of infected mosquitoes. 
West Nile virus (WNV) is transmitted to humans and mammals through the bites of infected mosquitoes around the globe \cite{doi:10.1126/science.1201010}.
It is a Flavivirus first identified in the West Nile region of Uganda in 1937 \cite{doi:10.1128/jvi.01963-10}. 
%After its first report, the cases of WNV has been reported in various parts of the world that includes Asia, Africa, Europe and North America \cite{doi:10.1126/science.1201010}.
WNV primarily infects only the birds act as a reservoir, but it can also infect other animals, including horse and humans \cite{doi:10.1128/cmr.00045-12}. 
The mammals act as a dead-end host and do not have an active role in disseminating WNV \cite{10.1093/auk/124.4.1121}. 
Most humans infected with WNV do not develop any symptoms, but in some cases it can create a severe neurological disease that can be life-threatening, especially in the elderly or immunosuppressed patients \cite{KRAMER2007171}.\par 
WNV was first detected in the United States in 1999 near New York and has since spread across the continental United States \cite{10.1093/jme/tjz151, refId0}.
After its introduction, WNV has become an endemic disease in the US, with seasonal outbreaks happening every year \cite{Allan}.
According the Centres for Disease Control and Prevention (CDC), more than 50,000 reported cases of WNV are being reported in the US since 1999 \cite{https://doi.org/10.1111/gcb.15842}.

%\textcolor{red}{Current situation in the States and the effort to eradicate the WNV}\\
%It is to be noted that the occurrence of WNF cases in the US has displayed significant fluctuations on yearly basis and no discernible long-term pattern can't be concluded from the limited dataset thus far \cite{doi:10.1289/ehp.0800487,  10.1093/jme/tjz151}. 
Occurrence of WNV cases in the US fluctuates significantly from year to year and no discernible long-term pattern in these fluctuations has been identified from the limited dataset thus far \cite{doi:10.1289/ehp.0800487,  10.1093/jme/tjz151}. 
WNV cases can be found throughout the contiguous $48$ states, the most affected areas with highest annual incidence are the parts of the Southwest, the Mississippi Delta region, the Great Plains, and the Rocky Mountain region \cite{10.1093/jme/tjz151, https://doi.org/10.1029/2022GH000708, refId0}.
WNV transmission by mosquitoes is heavily influenced by their blood-feeding behaviours, host-preferences and the climatic conditions in the environment. 
 
%\textcolor{red}{Feeding preference and its implications}\\
%The main vector of WNV is Culex pipiens mosquito and it is known to feed on a wide range of vertebrate hosts that includes birds, mammals and even humans \cite{doi:10.1098/rspb.2006.3575}.
A primary vector of WNV is the mosquito, \textit{Culex pipiens}. 
It is capable of feeding on a wide range of vertebrate hosts, including birds, mammals and even humans \cite{doi:10.1098/rspb.2006.3575}.
Mosquito feeding preferences strongly impact the intensity and timing of WNV infection in the United States.
A vector's “feeding index", or the proportion of blood meals obtained from a certain host species relative to it's abundance within the host community, provides a quantitative measure of a vector's preference for and reliance upon particular host species for blood meals \cite{doi:10.1098/rspb.2011.1282}.
A feeding index of $1$ signifies an opportunistic behaviour and a feeding index greater than $1$, implies a feeding preference for certain hosts \cite{doi:10.1098/rspb.2011.1282}.
By comparing the magnitude of this metric, we can gather information about the degree to which mosquitoes preferentially feed on certain hosts \cite{Levine, 10.1371/journal.pone.0039549}.
Exclusively avian feeding mosquito would boost the proliferation and dissemination of WNV in the ecosystem \cite{Sarah}.
A greater inclination to feed on mammals could lead to more instances of human cases of WNV \cite{HostSelection}.  
However, since mammals are the dead end hosts for WNV, a heightened focus on mammal feeding could perhaps ultimately lead to reduced viral amplification and environmental spread \cite{Komar}.
This information suggests that the feeding preferences play a pivotal role in the transmission of WNV.
%The feeding preference of mosquito is a highly influential factor and its impact on the intensity and timing of WNV infection in the United States has been documented  and the “feeding index” is an important metric utilised to assess the proportion of blood meals obtained from a certain host species relative to the abundance of that species within the host community \cite{doi:10.1098/rspb.2011.1282}.
%It can provide quantitative measures on the feeding preferences of vectors and their reliance on particular host species for blood meals \cite{Levine}.
%By comparing this metric, we can gather the information about the degree to which mosquitoes preferentially feed on certain hosts \cite{10.1371/journal.pone.0039549}.
Understanding the feeding-preferences of different mosquito species, though a deeper knowledge of the WNV transmission dynamics, can enable vector control to design improved abatement strategies \cite{doi:10.3920/978-90-8686-932-9_12}. 

%\textcolor{red}{Weather and how does it affect the dynamics of WNV?}\\
Weather plays an important role in influencing the dynamics of WNV.
%Certain weather conditions can influence the growth of mosquito populations, virus replication, transmission of the virus  and possibly feeding preferences too \cite{ROHR2011270, POH2019260}.
%Different weather factors such as temperature, rainfall, humidity, and wind patterns etc. can actively or passively affect the mosquito growth, abundance and their ability to transmit WNV \cite{10.7554/eLife.58511, Ruiz1, doi:10.1098/rsos.170017, 10.1371/journal.pone.0161510}. 
Temperature, rainfall, humidity and wind can actively or, passively influence mosquito growth and abundance, viral replication and viral transmission \cite{ROHR2011270, POH2019260,
10.7554/eLife.58511, Ruiz1, doi:10.1098/rsos.170017, 10.1371/journal.pone.0161510}.
Temperature has a profound impact on the dynamics of WNV \cite{10.7554/eLife.58511}. 
%Warm temperature generally helps faster mosquito development, reproduction and feeding activity. 
%This can yield an increase in mosquito population \cite{doi:10.1289/ehp.0800487}.  
Warmer temperatures generally increase mosquito development, reproduction, and feeding activity, leading to population growth \cite{doi:10.1289/ehp.0800487}.
Heightened temperature can reduce the incubation duration of the WNV within mosquitoes, thus facilitating the infected mosquitoes to transmit the WNV more rapidly \cite{10.1093/jmedent/43.2.309, 10.7554/eLife.58511}.  
Moreover, relatively high temperature can increase the replication and transmission of WNV within infected mosquitoes and it can potentially enhance the infectiousness of WNV infected mosquitoes \cite{10.7554/eLife.58511}. 
Geographic regions facing prolonged periods of high temperatures often experience higher WNV transmission and severe outbreaks \cite{10.1093/jmedent/43.2.309,  doi:10.1289/ehp.0800487}.  
However, the relationship between WNV and temperature is nonlinear as extreme heat and prolonged heatwaves can negatively impact the abundance of mosquito \cite{FAY2022147,  ROHR2011270}.
This may lead to more nuanced effects on the dissemination of WNV.
Rainfall and humidity play a significant role in shaping the dynamics of WNV by influencing the mosquito breeding sites \cite{doi:10.1098/rsos.170017}.
Adequate amounts of rainfall and humidity can provide a favourable conditions for mosquito reproduction and to provide breeding habitats to complete their life-cycle \cite{GIESEN2023100478}.
But heavy rainfall can have an adverse effect on the mosquito population as an excessive rainfall can flush away existing breeding sites and thus disrupting the growth of mosquito population \cite{10.1371/journal.pone.0001146, 10.1371/journal.pntd.0006935}. 
Furthermore, high humidity can regulate mosquito survival and activity \cite{https://doi.org/10.1111/ele.14228}.
Rainfall and humidity levels are also the guiding factors to determine the mosquito abundance and consequently controlling the potential spread of WNV in the local habitat. 
However, in our modelling effort, the model parameters are temperature dependent \cite{10.7554/eLife.58511, 10.1371/journal.pcbi.1006047, BHOWMICK2020110117, LAPERRIERE201199}.\par 
Mosquito management is a critical component of public health strategies with the aim to prevent mosquito-borne diseases such as malaria, dengue fever, and WNV.
It includes various methods, encompassing adulticide spraying, mosquito net distribution, and the elimination of breeding sites, such as standing water. 
This yields an immense effect on the growth of mosquito population and abundance \cite{10.1371/journal.pone.0246046}. 
Consequently, it also alters the WNV infection prevalence in a vector population \cite{10.1093/jme/tjad088}.\par

%\textcolor{red}{Previous modelling endeavours}\\
Public health policies developed both before and during an epidemic are often driven by lessons learnt from previous outbreaks.
Past outbreaks provide valuable insights into different features of the disease, its transmission dynamics, and the effectiveness of various control measures.
These experiences could potentially contribute valuable insights to the development of proper mathematical model that can help to prepare different response strategies, and preventive measures.%The authors in \cite{Vega} show that relative humidity plays an important role in the transmission of urban malaria and potentially other vector-borne epidemics 
The authors in \cite{Vega} show that  relative humidity, which is a pivotal factor influencing the spread of urban malaria and potentially other vector-borne epidemics, is often omitted from many mathematical models.
In the initial wave of the COVID-19 pandemic in 2020, lockdowns and quarantine measures are implemented, leading to notable changes in transmission patterns.
The lessons learnt from the early stages of the COVID-19 pandemic play a significant role in shaping current policy-making efforts \cite{Klump}.
Compartment-based models are particularly valuable in implementing such an outbreak, as they segment the population of concern into different groups based on their disease status.
This division allows the analytical calculation and to perform different  simulations to assess various control strategies or the progression of an outbreak \cite{Bergsman, BHOWMICK2023110213, BHOWMICK2023104827, BOWMAN20051107}. 
Additionally, several mathematical models are being constructed to investigate the transmission dynamics and the advancement of WNV infection \cite{10.1371/journal.pone.0227160,  EffectsofScaleonModelingWestNileVirusDiseaseRisk, doi:10.1289/EHP10287, 10.1371/journal.pntd.0010252}.
During a vector-borne disease outbreak, efficacy and the usage of different insecticide treatments to reduce the number of vectors are being examined in \cite{https://doi.org/10.1111/nrm.12165}.  
The authors in \cite{10.1371/journal.pone.0108452} find the basic reproduction number and utilised it to assess the abatement policies and conclude that the assumptions made in constructing the mathematical models are crucial and that different assumptions can lead to different epidemiological outcomes.
The feeding preferences of \textit{Culex Pipiens} were assessed while concentrating on their preferences for avian species as the primary host and the findings of the study acknowledge that the mosquitoes have a notable inclination to feed on the American robin compared to other hosts species \cite{doi:10.1098/rspb.2011.1282, doi:10.3920/978-90-8686-932-9_12}. 
The authors further add that the parameter describing this feeding preference is a key parameter that influencing the timing of the peak as well as the amplitude of WNV infection \cite{Levine, 10.1371/journal.pone.0039549, 10.3389/fevo.2022.993844}.
The authors in \cite{MALIK201860, https://doi.org/10.1111/nrm.12165, Demers}  investigate the effectiveness of insecticide based on surveillance data and the burden of WNV. 
The aforementioned work significantly ignores the inclusion of the influence of weather-driven factors as well as the stochasticity induced by the weather-driven parameters in the disease dynamics of WNV \cite{BHOWMICK2023110213, 10.1371/journal.pntd.0010252, doi:10.1289/EHP10287, LAPERRIERE201199, 10.1371/journal.pcbi.1006047}. 
Additionally, we know that the introduction of an epidemics and sustaining that, aren’t a single event phenomenon and the previous models fail to account for this aspect \cite{mbs:/content/journal/jgv/10.1099/vir.0.033829-0, Mann, doi:10.1098/rstb.2010.0054}.

%\textcolor{red}{our model and the questions we aspire to solve}
One of the main objectives of this study is to understand how the combination of temperature and feeding preferences of mosquito vectors across different field sites in Illinois can potentially shape the outbreaks of WNV. 
Additionally, we aim to determine the impact of continuous seasonal introduction of infected agents into our mechanistic, weather-driven Ordinary Differential Equations (ODE) based mathematical model.
Our work is organised in the following way: First we provide a description of our collected field data to be used for validating the model. 
%First we provide a detailed description of the field data we've gathered, which will be utilised for validating the model.
Second, we derive a deterministic, process based, climate driven ODE model that include the feeding preference of vector species and compute the analytic expression of basic reproduction number ($R_0$), after that we show the influence of different parameters on the transmission of WNV and the bifurcating nature of feeding preference. 
Finally, we validate our model with the trap data collected from the Cook County, Illinois, USA.  
%We developed a mechanistic, weather-driven Ordinary Differential Equation (ODE) based mathematical model to test the impacts of temperature, host preference, and vector abatement efforts on WNV transmission.

\section{Data collection}
\textbf{Mosquito Data}\par
A user agreement facilitated the acquisition of mosquito testing data from the Illinois Department of Public Health (IDPH) \cite{IDPH} for the period spanning from 2014 to 2018.
IDPH collects and consolidates data from local public health agencies and mosquito abatement districts throughout Illinois. 
They are responsible for a comprehensive statewide database that houses the results of mosquito testing for WNV.
To keep the consistency in mosquito collection and testing procedures across the state, the IDPH encourages use of the Centers for Disease Control and Prevention (CDC)-recommended a mosquito surveillance protocol. 
This protocol serves as a guideline that local health agencies and mosquito abatement districts are expected to adhere to, with the aim to standardise the processes involved in mosquito surveillance.
Typically, the local agencies employ gravid traps to collect mosquitoes. 
Subsequently, these agencies determine the sex and species of the captured mosquitoes. 
To test for the presence of WNV infection, they create pools comprising up to 50 mosquitoes of a single species from the samples obtained from each trap.
In cases where fewer than 50 mosquitoes are captured, the pool will contain the total number of mosquitoes collected. 

\begin{figure}[H]
\centering
\includegraphics[width=15cm]{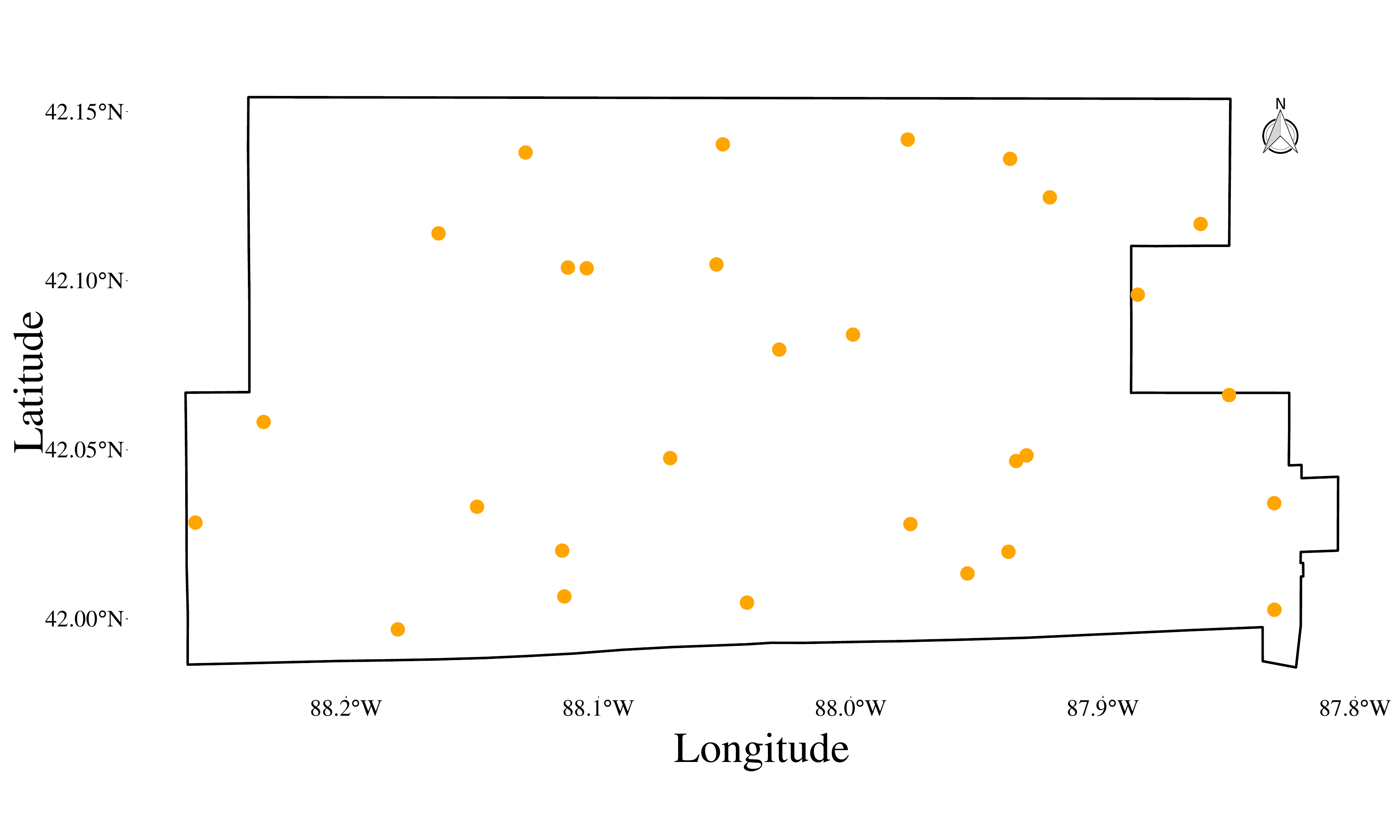}
\caption{Locations of mosquito trap data within Cook county, Illinois, USA, orange dots represent the locations of those trap stations.}
    \label{fig:Trap_Data}
\end{figure}

During the study period, common tests utilised to detect the presence of WNV in mosquitoes included antigen assays such as VecTest or the Rapid Analyte Measurement Platform (RAMP) test \cite{10.1371/journal.pone.0227160}.
In addition, certain pools underwent testing using Real-Time reverse transcriptase polymerase chain reaction (RT-PCR). 
If a pool is subjected to multiple types of tests, only the results retained from the RT-PCR test are considered for this analysis.
Throughout our analysis, we exclusively  rely on test results from pools of female \textit{Culex pipiens} mosquitoes. 
While not all mosquitoes are identified to species before testing, the majority of collected \textit{Culex pipiens} mosquitoes in this region are typically classified as either  \textit{Culex pipiens} or  \textit{Culex restuans}.
To  identify the positions of the mosquito traps, we  rely on the pre-existing latitude and longitude data stored in the IDPH database. 
Our analysis encompass all the trap locations documented within Cook County between 2014 and 2019.
Figure \ref{fig:Trap_Data}, shows mosquito trap locations within Cook County, Illinois, USA.

\textbf{Human Illness  Data}\par
Through a user agreement with the IDPH \cite{IDPH}, reported cases of human WNV in Illinois were acquired for this study. 
This research was ruled exempt by the University of Illinois Institutional Review Board (protocol $08686$) and was approved by the Illinois Department of Public Health Institutional Review Board (protocol $0950$).
All cases, both confirmed and probable, reported to the IDPH by medical and public health personnel within the study area are included. 
It is important to note that the state of Illinois mandates the reporting of WNV cases to local public health departments, which subsequently report all cases to the IDPH.
In the context of our research, probable cases of WNV are defined as those that exhibit clinical criteria consistent with the disease during the season when transmission is likely, in addition to meeting laboratory criteria through serology (IgM capture ELISA) or polymerase chain reaction (PCR) tests. 
Confirmed cases are those with definitive test results from either the IDPH or the CDC), providing confirmation of WNV infection.
All the human WNV cases in reported from 2017 to 2022 are aggregated for each year \cite{IDPH}.\par

\textbf{Bird Data}\par
A user agreement was established to enable the retrieval of bird testing data from the IDPH for the time period covering from 2017 to 2022 \cite{IDPH}.
The reported case counts of avian WNV mortality provided reflect the number of cases that are bing processed and completed by local health departments at the time of reporting.
The avian WNV fatality case counts may vary from those reported by the CDC due to differences in timing. 
%This discrepancy can occur because the reported counts reflect the data available at the time of reporting, which may differ from the CDC's data due to variations in reporting timelines.

\textbf{Weather Data}\par
\begin{figure}[H]
\centering
\includegraphics[width=8cm]{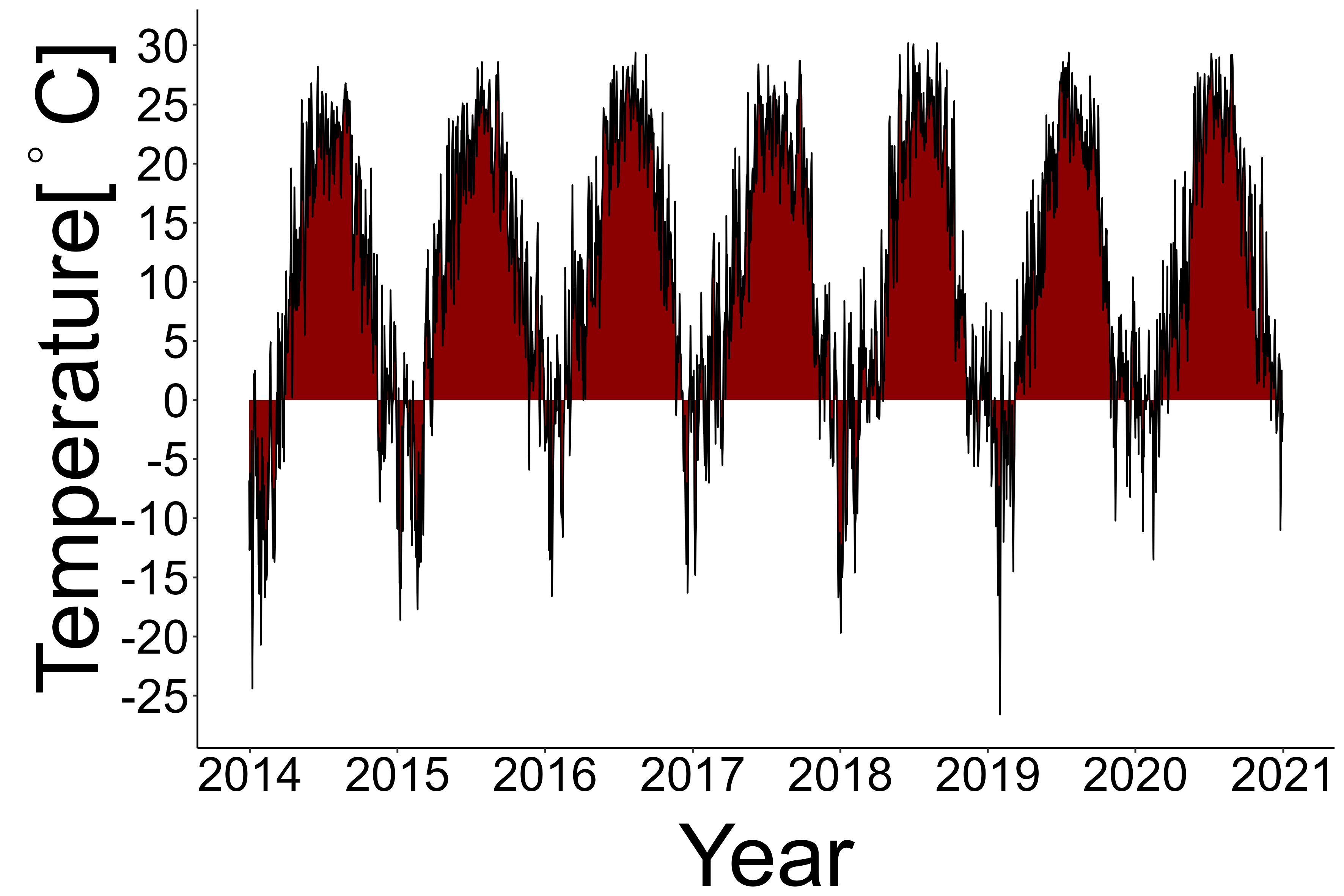}
\caption{Time series of daily mean temperature for the Cook county spanning from 2013 to 2022 provided by the PRISM Climate Group, affiliated with Oregon State University \cite{PRISMClimateGroup}.}
    \label{fig:Temp_Data}
\end{figure}

We obtain the spatial weather data on daily mean temperature from 2013 to 2022 from the PRISM Climate Group \cite{PRISMClimateGroup}.
Figure \ref{fig:Temp_Data} demonstrates the time-series of temperature data from 2013 to 2022.
The PRISM daily temperature data is obtainable as 4 km resolution spatial grids, computed through the utilisation of interpolation and statistical methods.
These techniques involve merging point data from weather monitoring networks nationwide with topographic data.

\section{Description of mathematical model}
\begin{figure}[H]
\centering
\includegraphics[width=8cm]{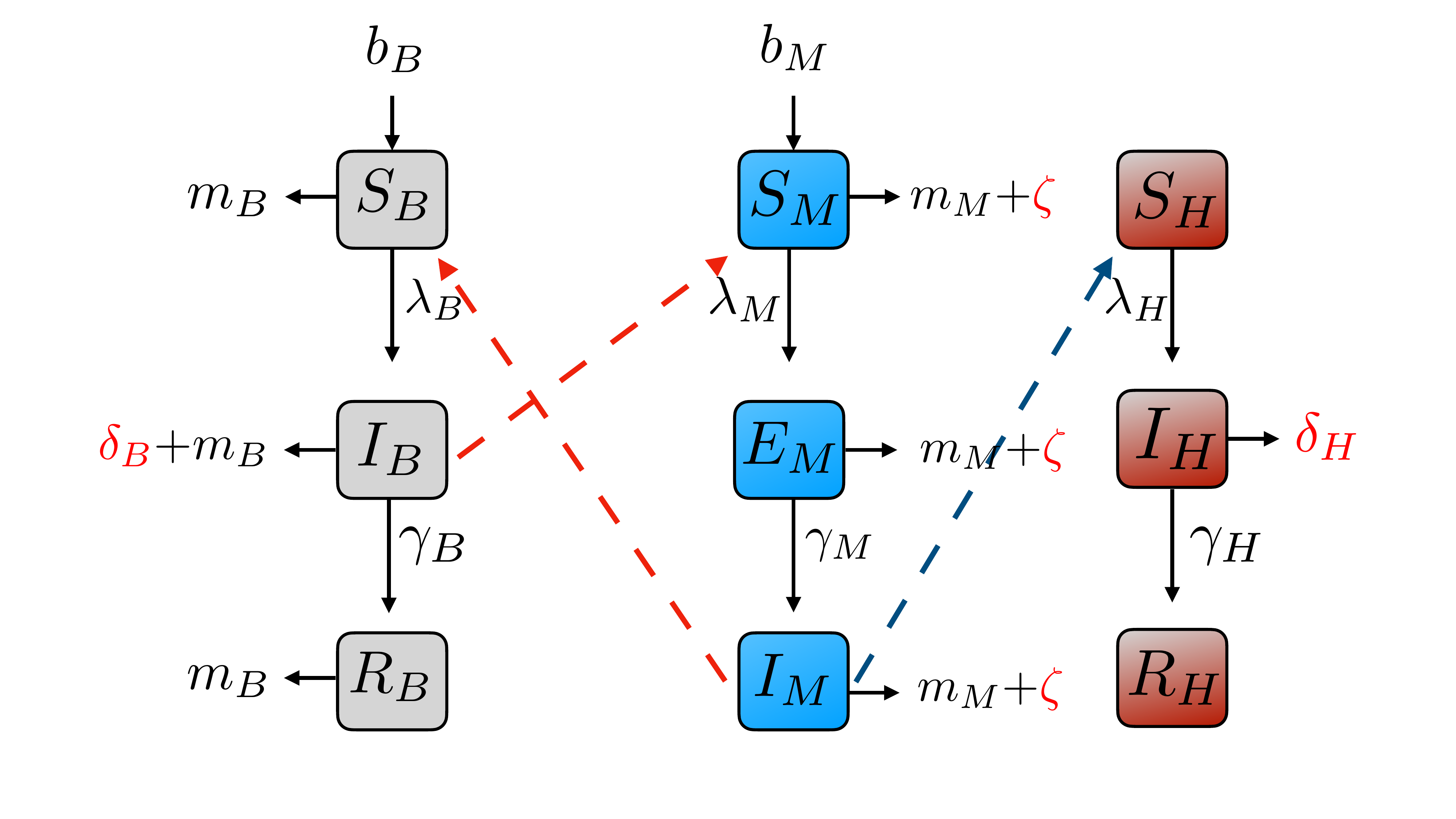}
\caption{ Flow diagram of WNV infection model. 
Infection cycle in mosquitoes (vector), birds (preferred host) and human  populations. 
The infection process involves the transmission from infected mosquito-to-bird and infected bird-to-mosquito (red dashed line) and infected mosquito-to-human (blue dashed line), demographic flux of individuals with different health status (black arrows). 
Mortality due to WNV and ULV spray are mentioned in red.
For the parameters description please see  Table \ref{table:1} and \ref{table:2}.}
    \label{fig:FlowChart}
\end{figure}

We construct a mechanistic ordinary differential equation (ODE) based an eco-epidemiological modelling framework like \cite{RUBEL2008166, Bergsman} of SEI-SIR type.
In our modelling effort, we  include avian hosts as the preferred host and human as the dead-end host to model zoonotic transmission.
We  consider a single mosquito vector species,  \textit{Culex pipiens} to model WNV transmission amongst the interacting species. 
Both primary and dead-end hosts are further divided into different compartments according to their health status as susceptible ($S_i$), infected ($I_i$) and recovered ($R_i$) and the vector species is divided into three compartments as susceptible ($S_M$), exposed ($E_M$) and infected ($I_M$), where $i = B, H$ and $B$ stands for avian and $H$ stands for human population. 
We include demographic process in mosquito and bird population but not in human population while constructing the ODE model.
A summary of the infection cycle is given in the Figure \ref{fig:FlowChart}.

The resulting model equations (\eqref{Eq1mos}, \eqref{Eq2bir}, \eqref{Eq3hum}) are given below in the following section \ref{Maintransmissionmodel}.

\subsubsection{Main transmission model}\label{Maintransmissionmodel}

Adult female mosquito model system (SEI Type) is described as follows
\par
\begin{eqnarray}\label{Eq1mos}
\frac{dS_M}{dt} &=& b_M(T)N_M-m_M(T)S_M-\lambda_M(T)S_M\\
\frac{dE_M}{dt} &=& \lambda_M(T)S_M -\gamma_M(T)E_M-m_M(T)E_M\\
\frac{dI_M}{dt} &=& \gamma_M(T)E_M-m(T)I_M
\end{eqnarray}

Bird model system  (SIR Type) is modelled as follows\par
\begin{eqnarray}\label{Eq2bir}
\frac{dS_B}{dt} &=& b_BN_B-m_BS_B-\lambda_B(T)S_B\\
\frac{dI_B}{dt} &=& \lambda_B(T)S_B -\gamma_BI_B-m_BI_B-\delta_BI_B\\
\frac{dR_B}{dt} &=& \gamma_BI_B-m_BR_B
\end{eqnarray}

and the human model system  (SIR Type) is described as \par
\begin{eqnarray}\label{Eq3hum}
\frac{dS_H}{dt} &=& -\lambda_H(T)S_H\\
\frac{dI_H}{dt} &=& \lambda_H(T)S_H -\gamma_HI_H-\delta_HI_H\\ %+poisson(\alpha I_M)
\frac{dR_H}{dt} &=& \gamma_HI_H
\end{eqnarray}

We include the hibernation of female adult mosquitoes (diapause) into our mathematical model \eqref{Eq1mos}, \eqref{Eq2bir} and \eqref{Eq3hum}, where the non-diapausing mosquito is defined by $\alpha_M$.
We also incorporate the mosquito-to-host ratio given by parameters $\phi_B$, $\phi_H$ and the forces of infection ($\lambda_i$, $i =M, B, H$) are defined by 
$\lambda_M(T) \rightarrow \frac{\alpha_M\beta_1 \eta(T)\alpha_F I_B}{\alpha_FN_B+N_H}$,
$\lambda_B(T) \rightarrow \frac{\phi_B\alpha_M\beta_2 \eta(T) \alpha_F I_M}{\alpha_FN_B+N_H}$,
$\lambda_H(T) \rightarrow \frac{\phi_H\alpha_M\beta_3 \eta(T)I_M}{\alpha_F N_B+N_H}$.
According to this way of formulation of the force of infection is a function of temperature dependent biting rate of mosquitoes $\eta(T)$ \cite{BHOWMICK2020110117, LAPERRIERE201199}.
Here, $\beta_1$ represents the transmission probability that an infected bird transmits WNV to a susceptible mosquito, $\beta_2$ means an infected mosquito transmits WNV to a susceptible bird, $\beta_3$ means an infected mosquito transmits WNV to a susceptible human, $\phi_B$ and $\phi_H$ represent mosquito-to-host ratio (bird and human, respectively)
Model parameters and their descriptions are given in the tables \ref{table:1} and \ref{table:2}.
%%%%%%%%%%%%%%%%%%%%%%%%%%%%%%%%%%%%%%%%
\begin{table}[H]
\centering
\begin{tabular}{||c c  ||} 
 \hline
 Variables & Definition  \\ [0.5ex] 
 \hline\hline
$S_M$ & Susceptible mosquito   \\ 
$E_M$ & Exposed mosquito   \\
$I_M$ & Infected mosquito  \\
$S_B$ & Susceptible birds   \\ 
$I_B$ & Infected birds   \\
$R_B$ & Recovered birds  \\
$S_H$ & Susceptible human   \\ 
$I_{HN}$ & Infected human with neuroinvasive disease\\
$I_{HNN}$ & Infected human with non-neuroinvasive disease\\
$R_H$ & Recovered human  \\
$N_A$ & Total population (A$=$ M, B, H)  \\
[1ex] 
 \hline
\end{tabular}
\caption{Model variables and their definitions.}
\label{table:1}
\end{table}

%%%%%%%%%%%%%%%%%%%%%%%%%%%%%%%%%%%%
\begin{table}[H]
\centering
\begin{tabular}{||c c  c ||} 
 \hline
 Parameters & Definition  & Values \\ [0.5ex] 
 \hline\hline
 $b_M(T)$ & \tiny{Mosquito birth rate}  & f(T) \cite{BHOWMICK2020110117, LAPERRIERE201199} \\ 
 $m_M(T)$ & \tiny{Mosquito mortality rate} & f(T) \cite{BHOWMICK2020110117, LAPERRIERE201199} \\
$\gamma_M(T)$ & \tiny{Incubation rate}  & f(T) \cite{BHOWMICK2020110117, LAPERRIERE201199}\\
$b_B$ & \tiny{Birds birth rate}  & 0.00342 \cite{BHOWMICK2020110117}\\
$m_B$ & \tiny{Bird mortality rate}  & 0.0012 \cite{BHOWMICK2020110117}\\ 
$\gamma_B$ & \tiny{Recovery rate of birds} & 0.182 \cite{BHOWMICK2020110117} \\
$\delta_B$ & \tiny{WNV induced death rate in bird}  & 0.26 \cite{BHOWMICK2020110117}\\
$\gamma_{HA}$ & \tiny{Recovery rate of human (A =N, NN)}  & 0.5 \cite{LAPERRIERE201199}\\
$\delta_H$ & \tiny{WNV induced death rate in neuroinvasive disease cases} & 0.004 \cite{LAPERRIERE201199} \\
$\lambda_i$ & \tiny{Force of infection ($i = M, B, H$)} & f(T) \cite{BHOWMICK2020110117, LAPERRIERE201199} \\
$\eta(T)$ & \tiny{Biting rate} & f(T) \cite{BHOWMICK2020110117, LAPERRIERE201199} \\
$\alpha_F$ & \tiny{Feeding index for birds } & $[5-40]$ \cite{doi:10.1098/rspb.2011.1282} \\
$\beta_1$ &\tiny{Transmission probability: $I_B \rightarrow S_M$}  & $[0, 1]$ \cite{BHOWMICK2020110117, LAPERRIERE201199}\\
$\beta_2$ & \tiny{Transmission probability: $I_M \rightarrow S_B$}  & $[0, 1]$ \cite{BHOWMICK2020110117, LAPERRIERE201199}\\
$\beta_3$ & \tiny{Transmission probability: $I_M \rightarrow S_H$} & $[0, 1]$  \cite{LAPERRIERE201199} \\
$p$ & \tiny{Fraction of WNV infected human will develop Neuroinvasive disease}& 0.006 Assumed\\
$\zeta_0$ & \tiny{ULV treatment effectiveness}& 0.5 \cite{10.1093/jme/tjad088}\\
$\phi_B$  & \tiny{Mosquito-to-bird ratio} & $[10-30]$ \cite{BHOWMICK2020110117}\\
$\phi_H$  & \tiny{Mosquito-to-human ratio} & 0.03 \cite{LAPERRIERE201199}\\
$\alpha_M$  & \tiny{Non-diapausing mosquitoes} & 0.5 \cite{ LAPERRIERE201199}\\
$\psi_B$  & \tiny{Infectious birds introduction rate} & $[10-130]$ Assumed\\
[1ex] 
 \hline
\end{tabular}
\caption{Model parameters and their definitions.}
\label{table:2}
\end{table}
\par
Now, suppose we classify the infected human cases into two categories (i) Neuroinvasive disease ($I_{HN}$) and (ii) non-neuroinvasive disease ($I_{HNN}$) cases then the model \eqref{Eq3hum} changes into the following:

\begin{eqnarray}\label{Eq34hum}
\frac{dS_H}{dt} &=& -\lambda_H(T)S_H\\
\frac{dI_{HN}}{dt} &=& p\lambda_H(T)S_H -\gamma_{HN}I_{HN}-\delta_HI_{HN}\\ %+poisson(\alpha I_M)
\frac{dI_{HNN}}{dt} &=& (1-p)\lambda_H(T)S_H -\gamma_{HNN}I_{HNN}\\
\frac{dR_H}{dt} &=& \gamma_{HN}I_{HN}+\gamma_{HNN}I_{HNN}
\end{eqnarray}

\subsubsection{Modified transmission model: applying adulticide}
Aerial applications of organophosphate or pyrethroid insecticides in ultra-low volume (ULV) are being proven to be effective in managing adult mosquito populations and possibly preventing outbreaks of WNV by reducing the number of vectors \cite{10.1371/journal.pone.0108452, Demers}.
Let us extend the model \eqref{Eq1mos} after including the usage of ULV (insecticide) and the model equation becomes
\begin{eqnarray}\label{Eq2mos}
\frac{dS_M}{dt} &=& b_M(T)N_M-(m_M(T)+\zeta)S_M-\lambda_M(T)S_M\\
\frac{dE_M}{dt} &=& \lambda_M(T)S_M -\gamma_M(T)E_M-(m_M(T)+\zeta)E_M\\
\frac{dI_M}{dt} &=& \gamma_M(T)E_M-(m(T)+\zeta)I_M
\end{eqnarray}

The functional form of $\zeta$ is important to formulate. 
It is a pulsating system as ULV is applied during the summer only \cite{10.1093/jme/tjad088, 10.1371/journal.pone.0108452}.
Let us take the simple step function to describe it after following \cite{10.1371/journal.pone.0108452}.
The following step function $\zeta(t)$ represents the mortality rate of mosquitoes due to the treatment of adulticide and it is a proportional reduction. 
\begin{equation}
\zeta (t) =
 \begin{cases} 
     \zeta_0 & t^{\mbox{apply}}\leq t\leq t^{\mbox{apply}}+ t^{\mbox{duration}} \\
      0 & \mbox{Otherwise} 
   \end{cases}
\end{equation}
Here $ t^{\mbox{apply}}$ is the day of the ULV treatment application,  $t^{\mbox{duration}}$ is the duration of the treatment and 
$\zeta_0$ is the the daily ULV treatment effectiveness.
\par

\subsubsection{Modified transmission model: continuous introduction of infected host}
\subparagraph{Introduction function}
The main WNV transmission model that we construct, employ a continuous flow of infected individuals into the infected compartment to better represent real introduction dynamics rather than a single introduction event which happens at a single fixed point at a time of simulation.
This phenomenon does not necessarily can capture the events of multiple introductions of infected agents into the infected compartments during multiple years \cite{Grub, doi:10.1098/rstb.2010.0054}.
So, after following the work in \cite{Grub, Pet}, we also model the introduction of WNV into the model system as follow:
\begin{eqnarray}\label{IntroFunct}
\psi_B(t) = A_0\left(\frac{e^{(A_m-t)}/A_w}{(1+e^{(A_m-t)}/A_w)^2}\right)
\end{eqnarray}
\begin{figure}[H]
\centering
\includegraphics[width=8cm]{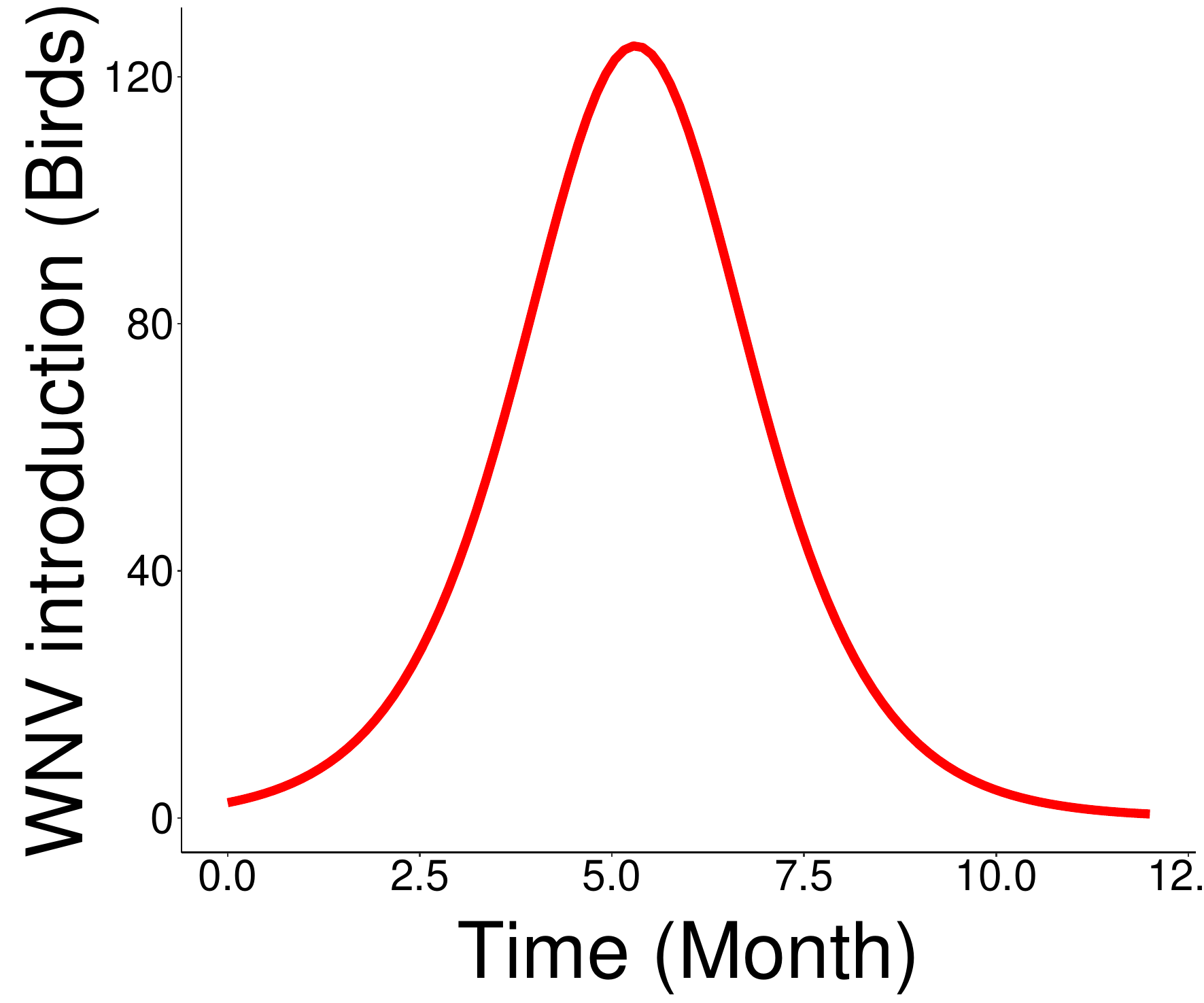}
\caption{Schematic of introduction function in WNV transmission model \eqref{Eq3bir}}
    \label{fig:WNVIntro}
\end{figure}
Here, $A_0$ represents the peak, $A_m$ represents midpoint, $A_w$ represents the width and $i_B(t)$ is the number of infected agents being introduced at time $t$.
So, after using the above mentioned introduction function \eqref{IntroFunct} into the model associated with the main host of WNV in \eqref{Eq2bir}, we get the following model
\begin{eqnarray}\label{Eq3bir}
\frac{dS_B}{dt} &=& b_BN_B-m_BS_B-\lambda_B(T)S_B\\
\frac{dI_B}{dt} &=& \lambda_B(T)S_B -\gamma_BI_B-m_BI_B-\delta_BI_B+I_B \psi_B(t)\\
\frac{dR_B}{dt} &=& \gamma_BI_B-m_BR_B
\end{eqnarray}

For a descriptive values of the model parameters, used in  \eqref{Eq1mos}, \eqref{Eq2mos},  \eqref{Eq2bir}, \eqref{Eq3bir}, \eqref{Eq3hum}, \eqref{Eq34hum}, please see the following articles \cite{10.7554/eLife.58511, 10.1371/journal.pcbi.1006047, BHOWMICK2020110117, LAPERRIERE201199, Pet, doi:10.1098/rspb.2011.1282}.
%\subsection{Mathematical Analysis}
\subsection{Basic Reproduction Number}
The Basic Reproduction Number ($R_0$) is an important metric in epidemiology as it determines the stability of the model system.
When $R_0 <1$, an outbreak will become extinct and if $R_0>1$, the disease will be established in the population. 
In this section we analytically derive the close form of $R_0$ after following \cite{doi:10.1098/rsif.2009.0386, VANDENDRIESSCHE200229}.
In our model system, $R_0$ represents the average number of secondary infected population, after the introduction of an infected agent into a completely susceptible interacting populations.
To construct the \textit{Next Generation Matrix}, we consider the equations associated with the spread of WNV from the model system \eqref{Eq2mos}, \eqref{Eq3hum} and  \eqref{Eq3bir}.
To apply the NGM approach \cite{doi:10.1098/rsif.2009.0386, VANDENDRIESSCHE200229} we consider the variables associated with WNV are $(E_M, I_M, I_B, I_{HN}, I_{HNN})$ and $(S_M, S_B, S_H)$ as the infectious-infected and  susceptible compartments.
Denoting $Y_I = (E_M, I_M, I_B, I_{HN}, I_{HNN})$ and $Y_S = (S_M, S_B, S_H)$, we can rewrite the associated system as the difference between the new-infection terms (inflow) and outflow terms then we have, 
\begin{equation}\label{R0Eqn1}
\frac{dY_I}{dt} =  \mathrm{F}(Y_S, Y_I)-\mathrm{V}(Y_S, Y_I)
\end{equation}
where
\begin{equation}\label{EqFmatrix}
\mathrm{F} =
\begin{bmatrix}
                              \frac{\alpha_M \beta_1\eta \alpha_F I_B S_M}{(\alpha_FN_B+N_H)}\\
                               0\\
                               \frac{\phi_B\alpha_M \beta_2\eta \alpha_F I_M S_B}{(\alpha_FN_B+N_H)}+\psi_BI_B\\
                               \frac{p\phi_H\alpha_M \beta_3\eta \alpha_F I_M S_H}{(\alpha_FN_B+N_H)}\\
			 \frac{(1-p)\phi_H\alpha_M \beta_3\eta \alpha_F I_M S_H}{(\alpha_FN_B+N_H)}
\end{bmatrix}
\end{equation}
and 
\begin{equation}\label{EqVmatrix}
\mathrm{V} =
\begin{bmatrix}
(\gamma_M+m_M+\zeta)E_M\\
-\gamma_ME_M+(m_M+\zeta)I_M\\
(m_B+\delta_B+\gamma_B)I_B\\
(\gamma_{HN}+\delta_H)I_{HN}\\
\gamma_{HNN}I_{HNN}
\end{bmatrix}
\end{equation}

So 
\begin{equation}\label{EqKLmatrix}
\mathrm{K_L} =
\begin{bmatrix}
0 & 0 & \frac{S_M^*\alpha_F\alpha_M\beta_1\eta}{(\alpha_FN_B^*+N_H^*)(\delta_B+\gamma_B+m_B)}& 0 & 0\\
0&0&0&0&0\\
\frac{S_B^*\alpha_F\alpha_M\beta_2\eta\gamma_M\phi_B}{(\alpha_FN_B^*+N_H^*)(m_M+\zeta)(\gamma_M+m_M+\zeta)}&\frac{S_B^*\alpha_F\alpha_M\beta_2\eta\phi_B}{(m_M+\zeta)(\alpha_FN_B^*+N_H^*)}&\frac{\psi_B}{\delta_B+\gamma_B+m_B}&0&0\\
\frac{S_H^*\alpha_M\beta_3\eta\gamma_M\phi_Hp}{(\alpha_FN_B^*+N_H^*)(m_M+\zeta)(\gamma_M+m_M+\zeta)}&\frac{S_H^*\alpha_M\beta_3\eta p\phi_H}{(m_M+\zeta)(\alpha_FN_B^*+N_H^*)}&0&0&0\\
\frac{S_H^*\alpha_M\beta_3\eta\gamma_M\phi_H(1-p)}{(\alpha_FN_B^*+N_H^*)(m_M+\zeta)(\gamma_M+m_M+\zeta)}&\frac{S_H^*\alpha_M\beta_3\eta (1-p)\phi_H}{(m_M+\zeta)(\alpha_FN_B^*+N_H^*)}&0&0&0
\end{bmatrix}
\end{equation}
 
%The matrix \eqref{EqKLmatrix}, can further be reduced to the following matrix
%
%\begin{equation}\label{EqKLmatrix}
%\mathrm{K_S} =
%\begin{bmatrix}
%\frac{SB^{*}\alpha_M\alpha_F\beta_2\eta\phi_B}{(N_B^{*}\alpha_F+N_H^{*})(\delta_B+\gamma_B+m_B)} & \frac{\psi_B}{\delta_B+\gamma_B+m_B}\\
%0 & \frac{S_M^{*}\alpha_F\alpha_M\beta_1\eta\gamma_M}{(N_B^{*}\alpha_F+N_H^{*})(m_M+\zeta)(m_M+\zeta+\gamma_M)}
%\end{bmatrix}
%\end{equation}
The spectrum radius is given by 
\begin{equation}\label{R0Model}
\rho(K_L) = \frac{1}{2}(h+\sqrt{h^2+4ag})
\end{equation}
where, 
$h = \frac{\psi_B}{\delta_B+\gamma_B+m_B}$, $a = \frac{S_M^*\alpha_F\alpha_M\beta_1\eta}{(\alpha_FN_B^*+N_H^*)(\delta_B+\gamma_B+m_B)}$ and $g = \frac{S_B^*\alpha_F\alpha_M\beta_2\eta\gamma_M\phi_B}{(\alpha_FN_B^*+N_H^*)(m_M+\zeta)(\gamma_M+m_M+\zeta)}$, $N_H^{*}$ and $N_B^{*}$ are the numbers of susceptible human and birds at the disease free equilibrium point.

The terms in the \eqref{R0Model} can also be biologically interpreted 
$\frac{\psi_B}{\delta_B+\gamma_B+m_B}$ represents the number of introduced infected birds per bird's infectious lifespan,
$\frac{S_B^*\alpha_F\phi_B}{\alpha_FN_B^*+N_H^*}$ is the is the number of initially susceptible preferred local birds per local host population at the disease-free equilibrium,
$\beta_2\eta \alpha_M$ describes the rate of successful transmission of WNV  through non-diapausing  mosquitoes to the preferred host,
 $\frac{\gamma_M}{\gamma_M+m_M+\zeta}$ depicts the proportion of mosquitoes that survive the incubation period under the influence of ULV, 
$\frac{1}{m_M+\zeta}$ is the lifespan of mosquito under the influence of ULV,
$\frac{S_M^*\alpha_M}{\alpha_FN_B^*+N_H^*}$ is the is the number of initially susceptible blood sucking mosquitoes per local host population at the disease-free equilibrium,
$\beta_1 \alpha_F\eta$ describes the rate of successful transmission of WNV to the preferred host ,
$\frac{1}{\delta_B+\gamma_B+m_B}$ represents the infectious lifespan of a bird, when $h= 0$, \eqref{R0Model}, reduces to $\sqrt{ag}$ and this simply can be interpreted as $\sqrt{(\text{Mosquito}\hookrightarrow \text{Bird})(\text{ Bird}\hookrightarrow \text{Mosquito})}$.\par
\section{Simulation}
\subsection{Impact of different parameters}
In this section we  explore the functional relationship among different parameters that can potentially influence the dynamics of WNV transmission. 
To find the functional relationship, we perform the simulations evaluated at 25 $\tccentigrade$ environmental temperature to include the temperature dependence of the vector population and varying two parameters at once and keeping the rest as fixed. 
The parameter values are mentioned in the Table \ref{table:2}.
To demonstrate the relationship, the number of susceptible mosquito and bird, total numbers of bird and human populations are set to $1000$, $100$ and $10000$, $1000$ respectively.  

\textbf{Relationship between $\phi_B$ and $\alpha_F$}\par
%By systematically varying the feeding index ($\alpha_F$) between $1$ and $35$ and analysing the resulting changes in transmission patterns, impact on the biological mechanisms we can gain insights into the role of $\alpha_F$ in the overall transmission dynamics of WNV.
%By methodically varying the magnitude of feeding index ($\alpha_F$) across a range from $1$ to $35$ and examining the consequent alterations in transmission patterns. 
%This way  we can glean valuable insights into how $\alpha_F$ influences the overarching transmission dynamics of WNV and its impact on the underlying biological mechanisms.
We vary the magnitude of the feeding index across a range from $1$ to $35$ to examine its impact on WNV transmission.
We know that a feeding index of $1$ signifies opportunistic feeding behaviour, whereas a feeding index exceeding $1$ indicates a preference for specific feeding.
In this way, we can understand the significance of either opportunistic feeding or, preference for specific feeding habit with mosquito-to-bird ratio. 
Here, we explore the dynamical relationship between mosquito-to-bird ratio ($\phi_B$) and the feeding index ($\alpha_F$) when $R_0$ is unity which is a critical value. 
When, $h = 0$, we have 
%\begin{flalign}\label{R0ModelpsiB}
\begin{equation}\label{R0whenhis0}
R_0{^2} = \left[\frac{S_M^*\alpha_F\alpha_M\beta_1\eta}{(\alpha_F N_B^*+N_H^*)(\delta_B+\gamma_B+m_B)}\right] \left[\frac{S_B^*\alpha_F\alpha_M\beta_2\eta\gamma_M\phi_B}{(\alpha_F N_B^*+N_H^*)(m_M+\zeta)(\gamma_M+m_M+\zeta)}\right]
\end{equation}
%\end{flalign}
%\eqref{R0ModelpsiB} yields the relationship between $\phi_B$ and $\alpha_F$ as the following
\begin{equation}\label{phibvsalphaF1}
\phi_B = \frac{R_0{^2}}{AB}\left( N_B^*+\frac{N_H^*}{\alpha_F}   \right)^2
\end{equation}
 where $A= \frac{S_M^*\alpha_M\beta_1\eta}{\delta_B+\gamma_B+m_B}$ and $ B= \frac{S_B^*\alpha_M\beta_2\eta\gamma_M}{(m_M+\zeta)(\gamma_M+m_M+\zeta)}$.
When $R_0 = 1$, at that critical value, \eqref{phibvsalphaF1} gives, 
\begin{equation}\label{phibvsalphaF2}
\phi_B^C = \frac{1}{A^CB^C}\left( N_B^*+\frac{N_H^*}{\alpha_F^C}   \right)^2
\end{equation}\par
\begin{figure}[H]
\centering
\includegraphics[width=8cm]{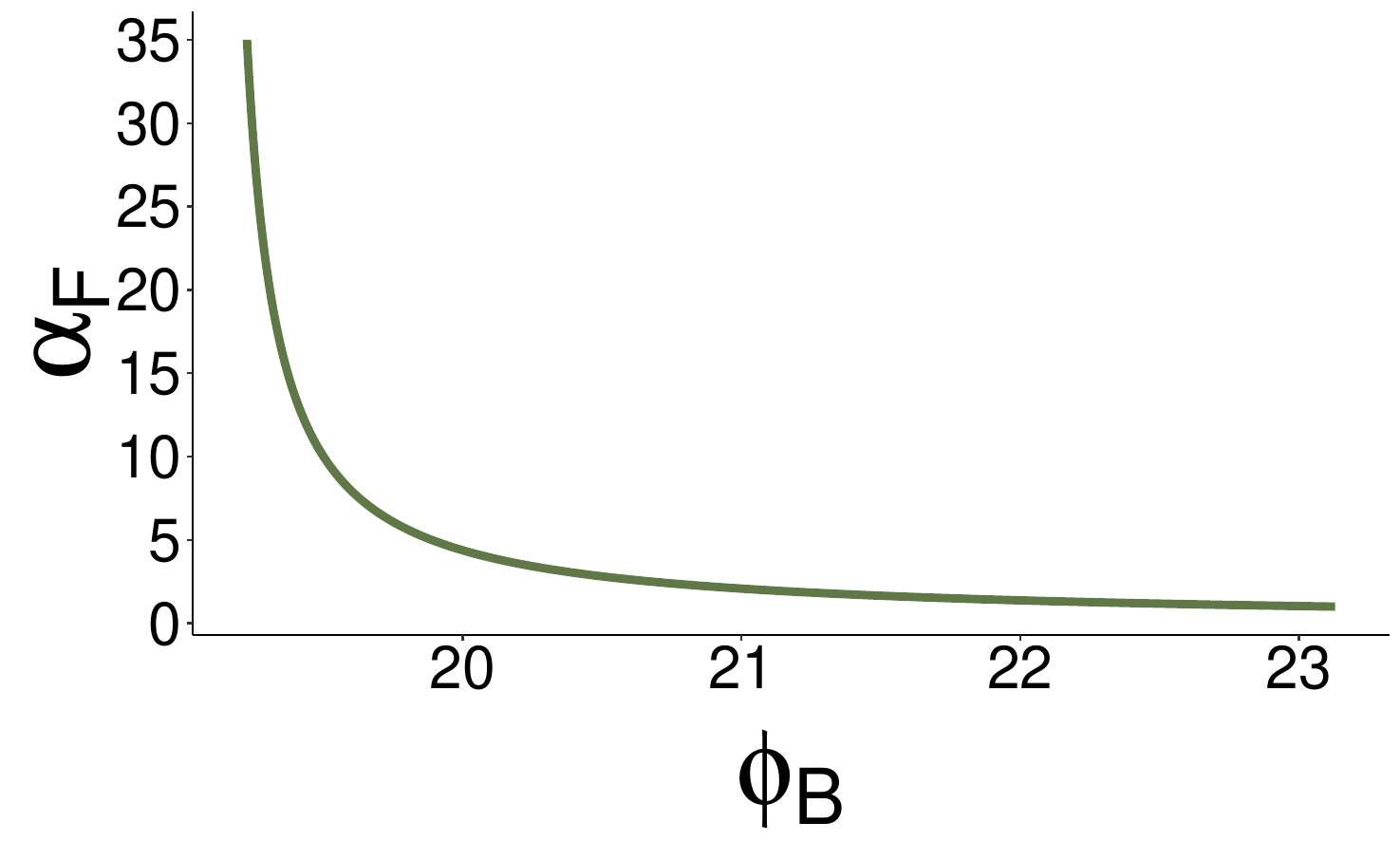}
\caption{The functional relationship between vector-to-host ratio ($\phi_B$) and feeding index ($\alpha_F$) according to \eqref{phibvsalphaF2} when the value of $R_0$ is $1$.}
    \label{fig:alphaFVsphiB}
\end{figure}
\begin{figure}[H]
\centering
\includegraphics[width=8cm]{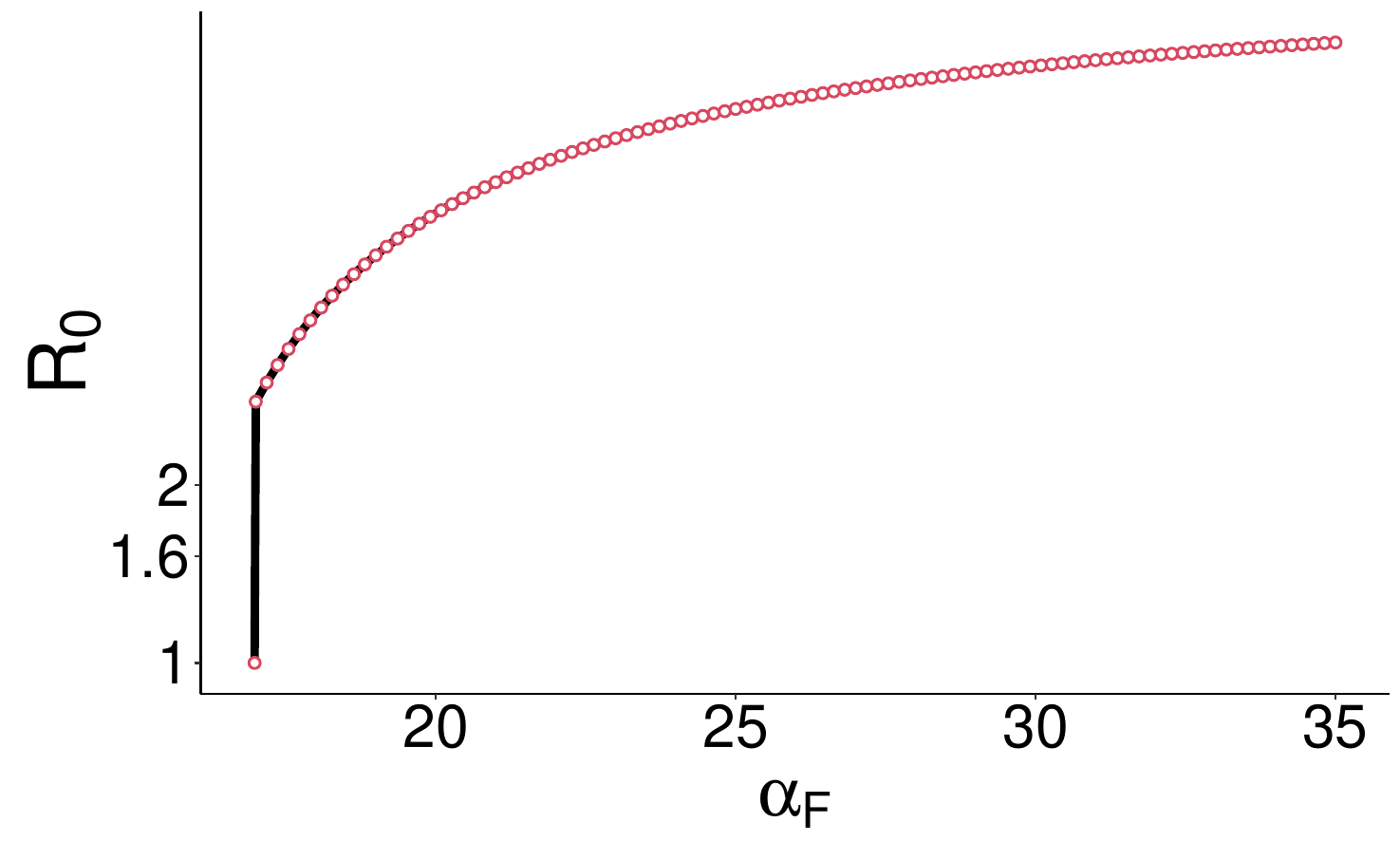}
\caption{Model predicted relationship between $R_0$ and $\alpha_F$}
    \label{fig:alphaVsR0}
\end{figure}

Figure \ref{fig:alphaFVsphiB} depicts the the  relationship between the feeding index and vector-to-host ratio according to \eqref{phibvsalphaF2}.
It is noticeable from the Figure \ref{fig:alphaFVsphiB} that $\phi_B$ and $\alpha_F$ follow the reciprocal squared functional dependence and this way we are able to depict a tractable trajectory between $\phi_B$ and $\alpha_F$. 
We can also observe that with the higher value of $\phi_B$, the value of $\alpha_F$, reduces and the dynamics changes when the magnitude of $\alpha_F$ lies between $4$ and $7$.
This implies that with the lower value to mosquto-to-bird ratio, the proportion of avian blood meals obtained by the mosquitoes increase in a nonlinear way and with the higher ratio between mosquito and bird, the feeding index approximately saturates. 
 This observation can lead us to conclude that $\alpha_F$ has an immense impact on the transmission dynamics.
 % and the moderate value of $\alpha_F$ can alter the WNV transmission dynamics. 
 To maintain an epidemic, a shift in the mosquito-to-bird ratio below a value of $20$ necessitates an adjustment of $\alpha_F$ from $4$ to $7$.
 In the Figure \ref{fig:alphaVsR0}, we further find the critical value of $\alpha_F$.

\textbf{Relationship between $\eta$ and $\alpha_F$} \par
We investigate the correlation between the rate of biting ($\eta$) and the feeding index ($\alpha_F$) when $R_0$ is unity.
From \eqref{R0whenhis0}, we have 
\begin{eqnarray}\label{alphaFvsEta}
R_0^2 &=& \left[\frac{A_1 \alpha_F \eta}{A_2(\alpha_F N_B^*+N_H^*)}\right] \left[ \frac{A_3 \alpha_F \eta}{A_4(\alpha_F N_B^*+N_H^*)} \right]\\ 
\eta^C &=& \sqrt{\frac{A_2A_4}{A_1A_3}} \left(N_B^*+N_H^*/\alpha_F^C\right),
\end{eqnarray}
when $R_0 = 1$ and $A_1 = S_M^*\alpha_M\beta_1$, $\delta_B+m_B+\gamma_B$, $A_3 = S_B^* \alpha_M\beta_2\gamma_M\phi_B$ and $A_4 =(m_M+\zeta)(m_M+\gamma_M+\zeta)$.

\begin{figure}[H]
\centering
\includegraphics[width=8cm]{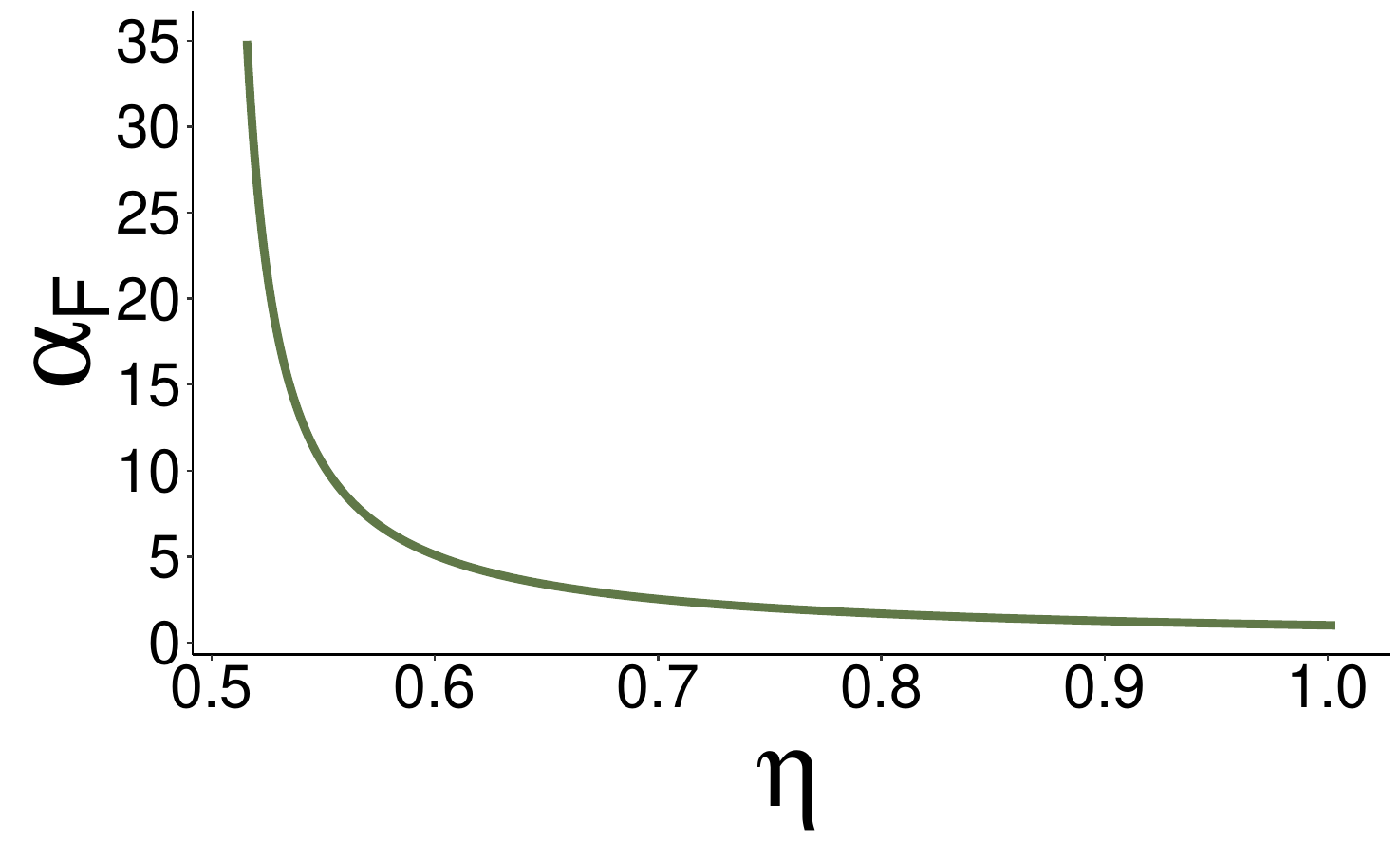}
\caption{Relationship between the biting rate ($\eta$) and the feeding index ($\alpha_F$) as described in \eqref{alphaFvsEta} when $R_0$ is $1$.}
    \label{fig:alphaVseta}
\end{figure}
Figure \ref{fig:alphaVseta} describes the functional relationship between the biting rate and the feeding index, and we can notice that it is also nonlinear. 
%A change in the moderate value of $\alpha_F$ (between $4$ and $7$), necessitates to maintain an epidemic outbreak.
%has the potential to change the dynamics of WNV transmission.
%the dynamics changes significantly. 
%We observe an interesting feature that with the increase in feeding index and biting rate reduces.
We notice that as the mosquito biting rate declines from a value of $1$ to around $0.55$, the feeding index must increase to maintain an $R_0$ of $1$.  
At biting rates of less than $0.55$, the feeding index must shift higher by an order of magnitude to maintain the outbreak.
%This can be attributed to the fact that with a higher biting rate, does not always yield higher feeding index and it possibly indicates that with the increase in biting does not necessarily result in more efficient feeding and hence not a significant increase in the magnitude of feeding index. 
%Other factors, such as resource availability, competition, or individual characteristics, can also possibly influence the feeding efficiency independently of the biting rate.  
%It's important to observe that the relationship between biting rate and feeding index in Culex pipiens can be complex.
%It may also depend on the local conditions, host availability, and other factors specific to the study area. 
%Further research and observations are necessary to gain a comprehensive understanding of this relationship in different contexts.

\textbf{Relationship between $R_0$ and $\alpha_F$} \par
We derive analytically the relation between $R_0$ and $\alpha_F$ from \eqref{R0whenhis0}. 
The relationship between $R_0$ and $\alpha_F$ is defined as 
\begin{equation}\label{alphaFvsR0}
R_0^2 = \sqrt{A_1A_2}\left[\frac{\alpha_F}{N_B^*\alpha_F+N_H^*}\right]
\end{equation}
where, $A_1 = \frac{S_M^*\alpha_M\beta_1\eta}{(\delta_B+\gamma_B+m_B)}$ and $A_2 = \frac{S_B^*\alpha_M\beta_2\eta\gamma_M\phi_B}{(m_M+\zeta)(\gamma_M+m_M+\zeta)}$.
Figure \ref{fig:alphaVsR0}, depicts the relationship between the basic reproduction number and the feeding index.
It the value of $\alpha_F$ is less than 6, then we do not observe any potential outbreak but when the value of $\alpha_F$ crosses its critical value, there is an outbreak as the value of $R_0$ is greater than $1$.

\textbf{Relationship between $R_0$ and $\zeta$} \par
\begin{figure}[H]
\centering
\includegraphics[width=8cm]{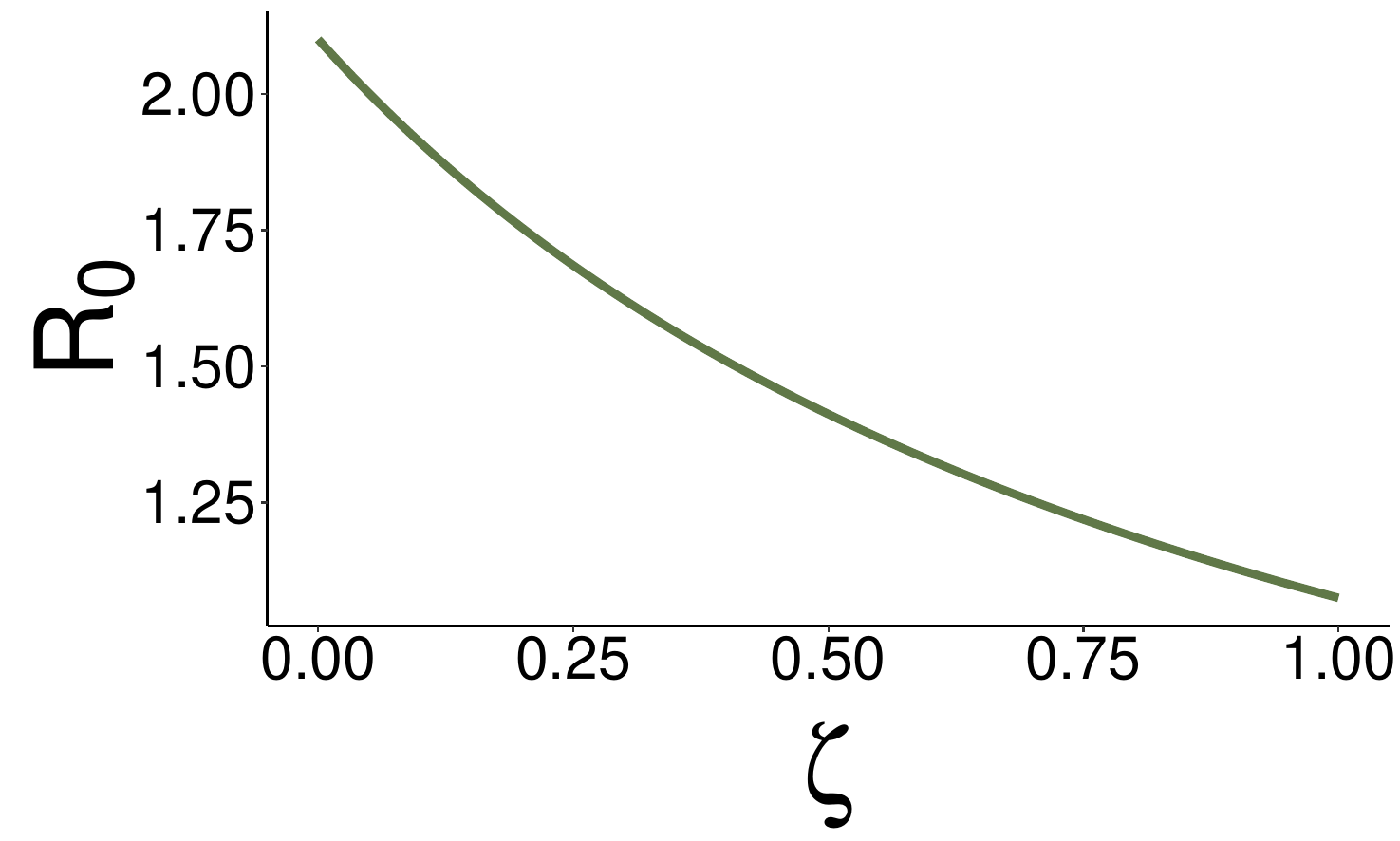}
\caption{Effect of $\zeta$ on $R_0$ as described in  \eqref{Eq1mos}}
    \label{fig:R0vsZeta}
\end{figure}

Here, we graphically explore the relationship between $R_0$ and $\zeta$ in the Figure \ref{fig:R0vsZeta} with the same parameter set as we  mention earlier. 
The Figure \ref{fig:R0vsZeta} is important from the perspective mosquito abatement and WNV control.
We can observe that with the available parameter values from the literature [citations], even after exhausting the resources, the magnitude of $R_0$ is not less than the unity.

%\begin{figure}[H]
%\centering
%\includegraphics[width=8cm]{Figures/ContourPlotEtavsAlphaFvsZetaR0-11.png}
%\caption{Effect of $\zeta$ on the relationship between $\alpha_F$ and $\eta$, when $R_0 = 1$ as described in  \eqref{R0whenhis0}}
%\label{fig:contourR0vsZeta}
%\end{figure}

%\begin{figure}[H]
%\centering
%\includegraphics[width=8cm]{Figures/alphaFvsIB.pdf}
%\caption{Effect of $\alpha_F$ on $I_B$}
% \label{fig:alphaFvsIB}
%\end{figure}
%
%\begin{figure}[H]
%\centering
%\includegraphics[width=8cm]{Figures/etavsIB.pdf}
%\caption{Effect of $\eta$ on $I_B$}
% \label{fig:etavsIB}
%\end{figure}
\textbf{Impact of $\alpha_F$ and $\eta$ on $I_B$} \par
In this set of simulations, we experiment the influence of varying the feeding index ($\alpha_F$) and the biting rate ($\eta$) between $0$-$30$ and $0$-$1$.
We perform the simulations only for a single season.
\begin{figure}[H]
    \centering
    \subfloat[\centering Effect of $\alpha_F$ on $I_B$]{{\includegraphics[width=6cm]{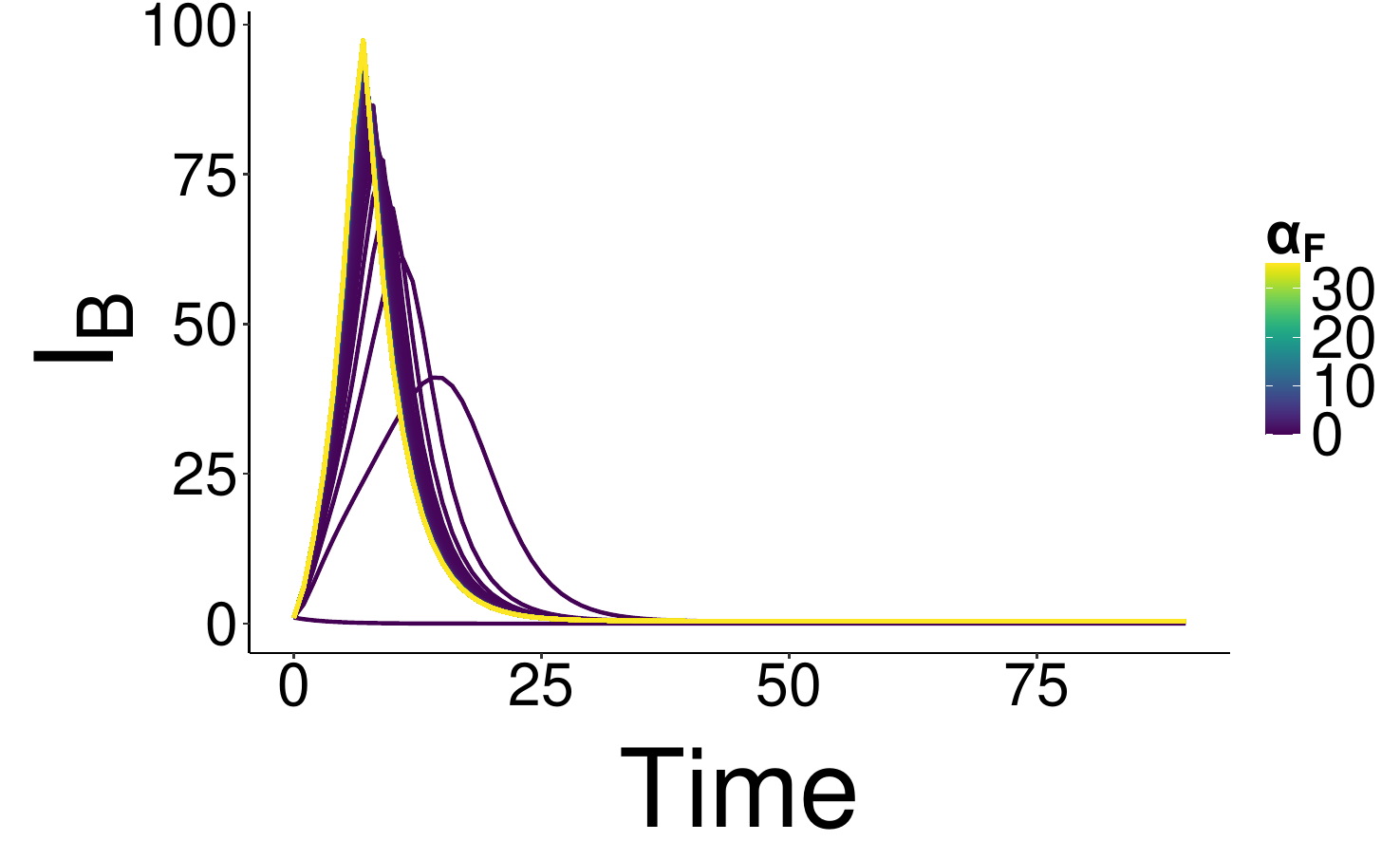} }}
    \label{fig:alphaFetavsIBa}%
    \qquad
    \subfloat[\centering Effect of $\eta$ on $I_B$]{{\includegraphics[width=6cm]{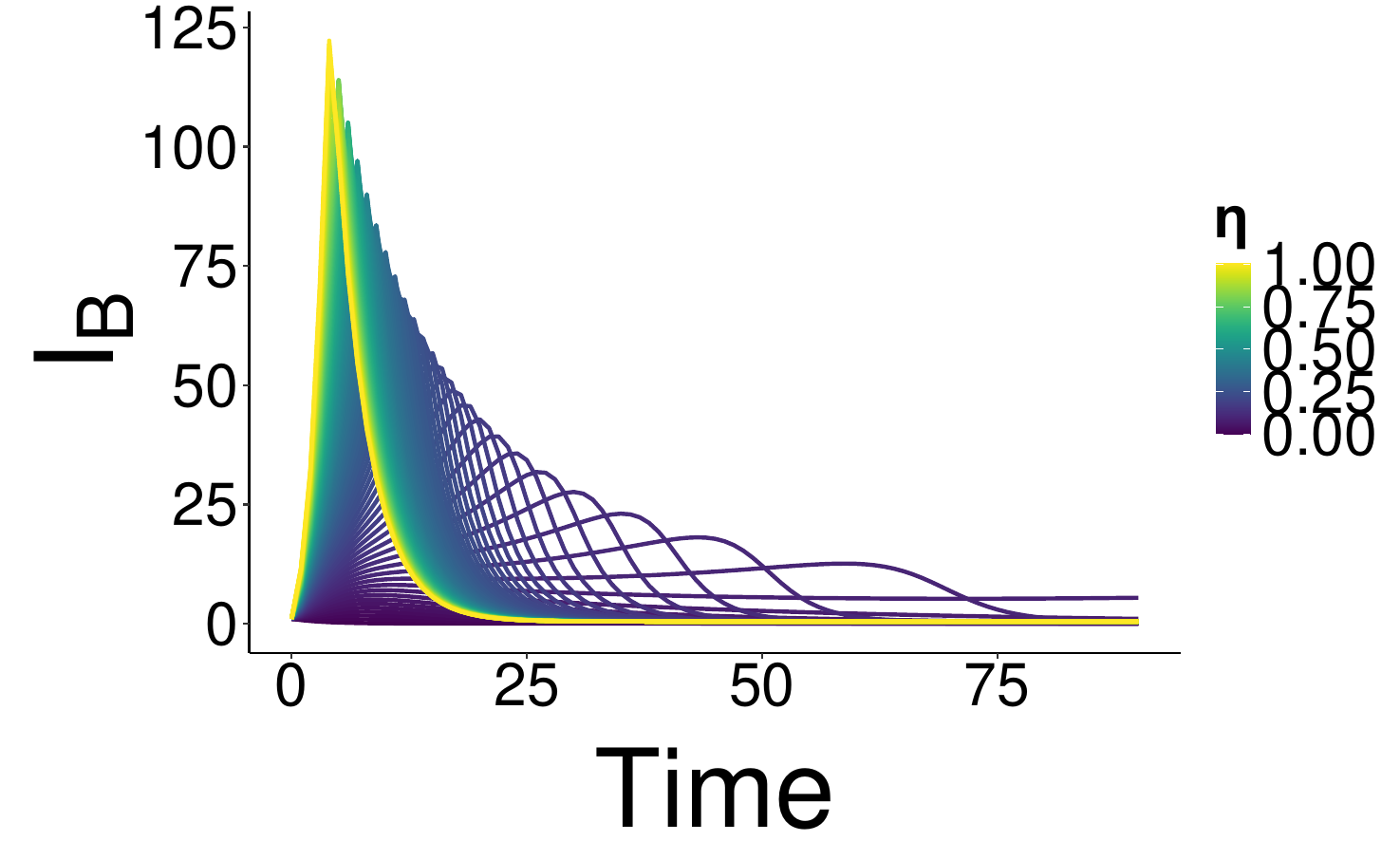} }}\label{fig:alphaFetavsIBb}%
    \caption{Model prediction of impact of varying feeding index ($\alpha_F$) and biting rate ($\eta$) on the infected bird population ($I_B$). }%
    \label{fig:alphaFetavsIB}%
\end{figure}

%\begin{figure}[H]
% \begin{subfigure}{0.1\linewidth}
%  \centering
%  \includegraphics{Figures/alphaFvsIB.pdf}
%  \caption{Effect of $\alpha_F$ on $I_B$}\label{fig:alphaFvsIB}
% \end{subfigure}%
%%
% \begin{subfigure}{0.1\linewidth}
%  \centering
%  \includegraphics{Figures/etavsIB.pdf}
%  \caption{Effect of $\eta$ on $I_B$}\label{fig:etavsIB}
% \end{subfigure}%
%\caption{Model prediction of impact of varying feeding index ($\alpha_F$) and biting rate ($\eta$) on the infected bird population ($I_B$).}
%\label{fig:IBvsalphaFandeta}
%\end{figure}

The Figures \ref{fig:alphaFetavsIB} depict the influence of $\alpha_F$ and $\eta$ on the number of infected birds.
In Figure \ref{fig:alphaFetavsIB} (a), we observe the influence of feeding index on the infection profile of the birds.
The peak number of infected birds in an outbreak increases with a higher feeding index.
%A higher feeding index accentuates an outbreak of WNV quicker compared to the situation with a lower magnitude of $\alpha_F$.
With a lower $\alpha_F$, an outbreak takes a while and the amplitude is relatively lower.
So, we can surmise that in a habitat patch, where the magnitude of $\alpha_F$ is higher, the propensity of an outbreak of WNV is relatively higher compared to another habitat patch where $\alpha_F$ is lower.
In Figure \ref{fig:alphaFetavsIB} (b), we show the impact of biting rate on the infected bird population.
Here we also observe that the peak number of infected birds increases with a higher biting rate.

\begin{figure}[H]
    \centering
    \subfloat[\centering Contour plot of $\alpha_F$, $\eta$ and $\zeta$]{{\includegraphics[width=6cm]{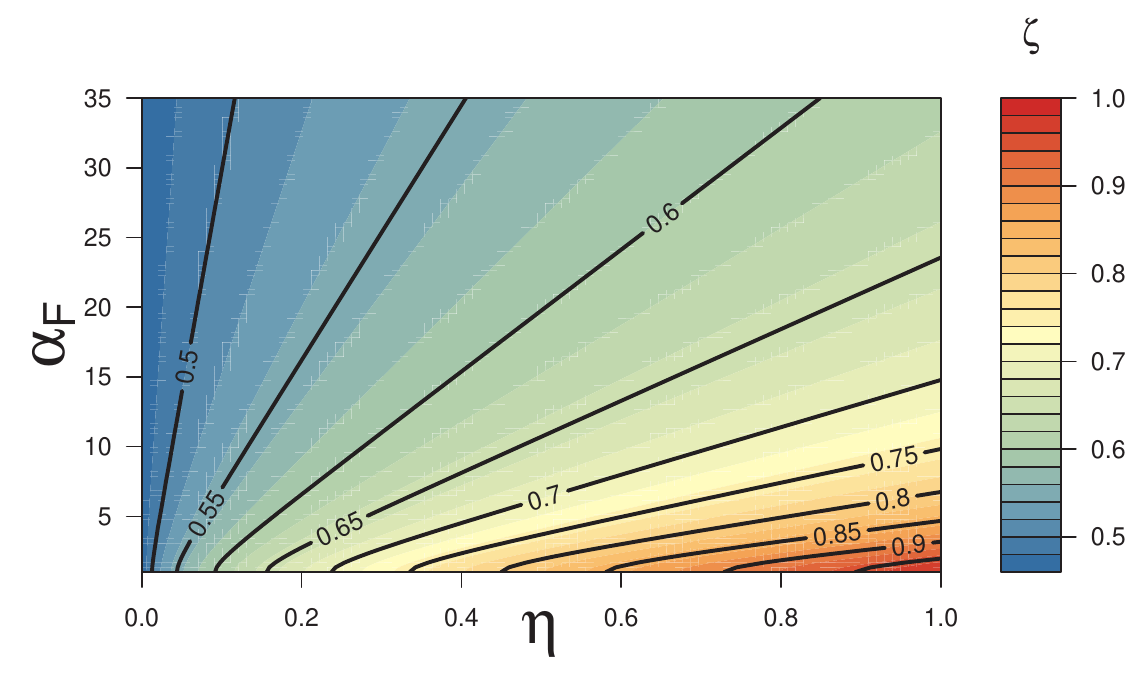} }}%
    \qquad
    \subfloat[\centering Contour plot of $\alpha_F$, $\zeta$ and $R_0$]{{\includegraphics[width=6cm]{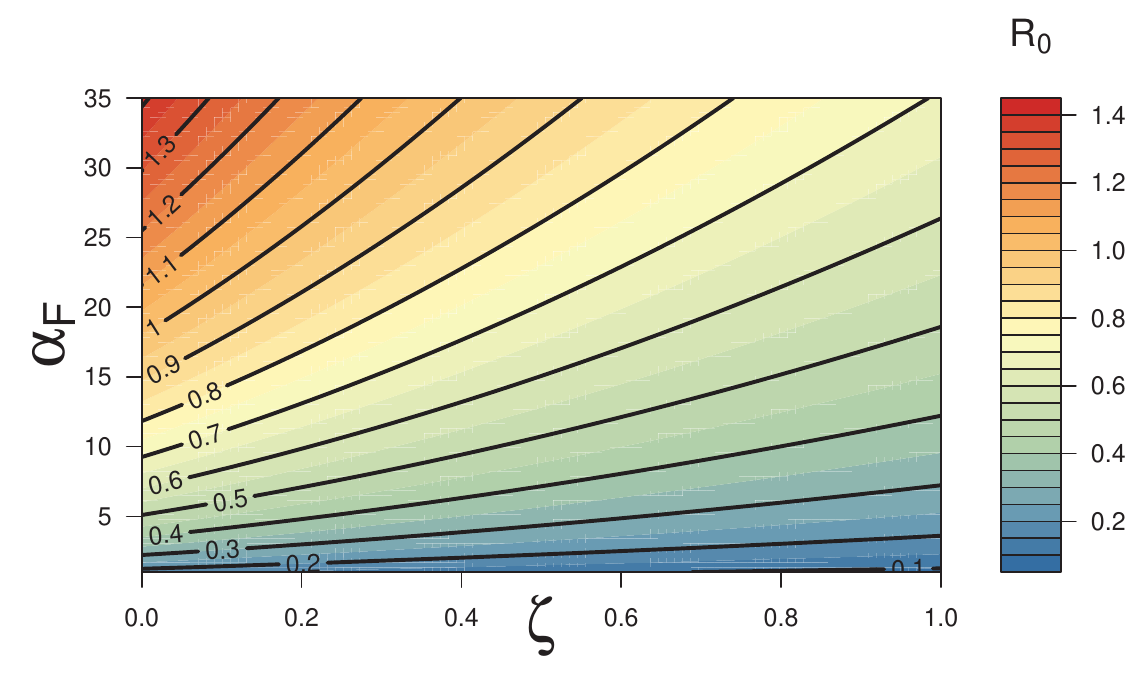} }}%
    \caption{(a) Contour plot of $\zeta$ as a function of feeding index ($\alpha_F$) and biting rate ($\eta$) and 
                 (b) Contour plot of $R_0$ as a function of feeding index ($\alpha_F$) and $\zeta$.
                 Parameter values used are as provided in Table \ref{table:2}
                 }%
    \label{fig:AllContourPlots}%
\end{figure}

%\begin{figure}[H]
%\centering
%\includegraphics[width=6cm]{Figures/contourplotR0vsalphafvszeta2.pdf}
%\caption{The relationship amongst $\zeta$, $\alpha_F$ ans $R_0$ as described in \eqref{R0Model}}
%\label{fig:contourR0vsZeta}
%\end{figure}
To demonstrate the impact of different parameters on the transmission of WNV, we  plot contour diagrams (Figure \ref{fig:AllContourPlots} (a) and (b)) while varying two parameters at a time.
Controlling the essential parameters, these diagrams can have a vital impact on managing the infected population.
To show the interplay of feeding index ($\alpha_F$) and biting rate ($\eta$) with the rate of adulticide spray ($\zeta$), we show a contour plot in Figure \ref{fig:AllContourPlots} (a).
Similarly, to show the impact of feeding index ($\alpha_F$) and the efficacy of adulticide ($\zeta$) on the basic reproduction number ($R_0$), we present a contour plot of $R_0$ as a function of $\alpha_F$ and $\zeta$ in Figure \ref{fig:AllContourPlots} (b).

Figure \ref{fig:AllContourPlots} (a) depicts the nonlinear functional relationship amongst $\alpha_F$, $\eta$ and $\zeta$ and it is also interesting to notice that the dependency between $\eta$ and $\zeta$ in the presence of heterogeneous values of feeding index is not merely a linear one. 
This possibly demonstrates the complexity of the model and the dynamics of WNV transmission cycle.
Figure \ref{fig:AllContourPlots} (b) demonstrates the impact of $\alpha_F$ and $\zeta$ on the magnitude of $R_0$.
With a lower magnitude of $\zeta$ and $\alpha_F$, we can observe that the value of $R_0$ is low but it is interesting to notice that when the value of $\alpha_F$ crosses the threshold value around $6$, the dynamics of the transmission cycle undergoes a change. 
Moreover, even if with a higher value of $\zeta$, the gradual increase in the magnitude of $R_0$ is noticeable. 
From the figure, it can be deduced that to reduce the burden of WNV, the magnitude of $\alpha_F$ should be low but whence it crosses the threshold value, it is difficult to reduce the value of $R_0$.
This is very similar to what we acknowledge in the  Figure \ref{fig:R0vsZeta}.

\subsection{Sensitivity Analysis}
We  perform sensitivity analysis to understand the behaviour of our mathematical model and identify the potential key transmission parameters that will influence $R_0$.
We utilise Latin Hypercube Sampling (LHS) to draw the sample values for each model parameter from their respective probability distributions.
We run the simulations 1000 times after employing the sampled parameter values to produce sets of model output and in our case the model response is the value of $R_0$.
This sampling technique generates input and output distributions that are beneficial in model assessment and to quantify the model parameter uncertainties.  
 
\begin{figure}[H]
\centering
\includegraphics[width=8cm]{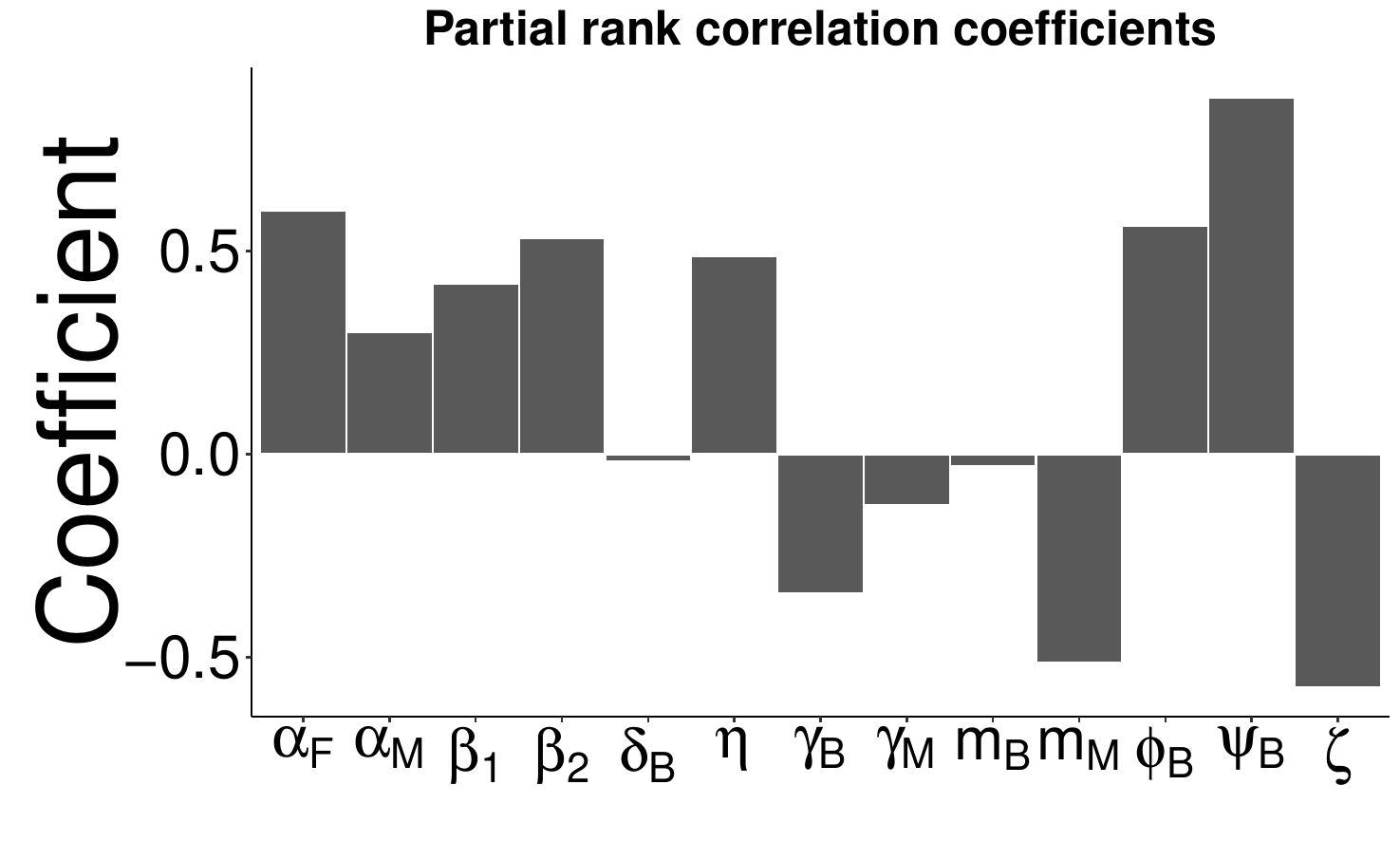}
\caption{PRCC sensitivity analysis of WNV $R_0$ as computed in \eqref{R0Model}}
    \label{fig:PRCC}
\end{figure}
The partial rank correlation coefficient (PRCC) is a widely utilised tool to conduct model parameter sensitivity and it quantifies PRCC values for each input parameters combination and model response variables.
The PRCC rankings reveal the parameters that strongly influence model outcomes and this helps to identify the important biological mechanisms that define the dynamics and the course of pathogen transmission.
We use the sensitivity package \cite{sensivity}, and for the LHS scheme we utilise the lhs package \cite{lhs} in R \cite{R}.
In Figure \ref{fig:PRCC} we present the PRCCs for the most significant parameters. 
We observe that the mortality rate of mosquito ($m_M$), efficacy of adulticide ($\zeta$) show a strong negative correlation with $R_0$.
Biting rate ($\eta$),  effective transmission rate from infected mosquito to susceptible bird ($\beta_2$), non-diapausing mosquito ($\alpha_M$), the rate of introduced infected hosts ($\psi_B$), feeding index ($\alpha_F$), mosquito-to-bird ratio ($\phi_B$) and effective transmission rate from infected birds to susceptible mosquitoes ($\beta_1$) show strong positive correlations with the model output, whereas the mortality rate of infected bird ($m_B$), incubation rate in mosquitoes and birds ($\gamma_M$, $\gamma_B$), WNV induced death rate ($\delta_B$) are not so significant.\par
Therefore, we conclude that the parameters with the strongest influence on the basic reproduction number are mortality rate of mosquitoes ($m_M$), efficacy of adulticide ($\zeta$), rate of introduced infected agents ($\psi_B$), feeding index ($\alpha_F$), mosquito-to-bird ratio ($\phi_B$) and the biting rate.

\subsection {Model Validation}
It is equally necessary to replicate the reported or, observed data through a constructed mathematical model to ensure the model's suitability and feasibility for making reliable predictions and inferences about the underlying system.
It can also help to identify model deficiencies and guide improvements.

\begin{figure}[H]
\centering
\includegraphics[width=8cm]{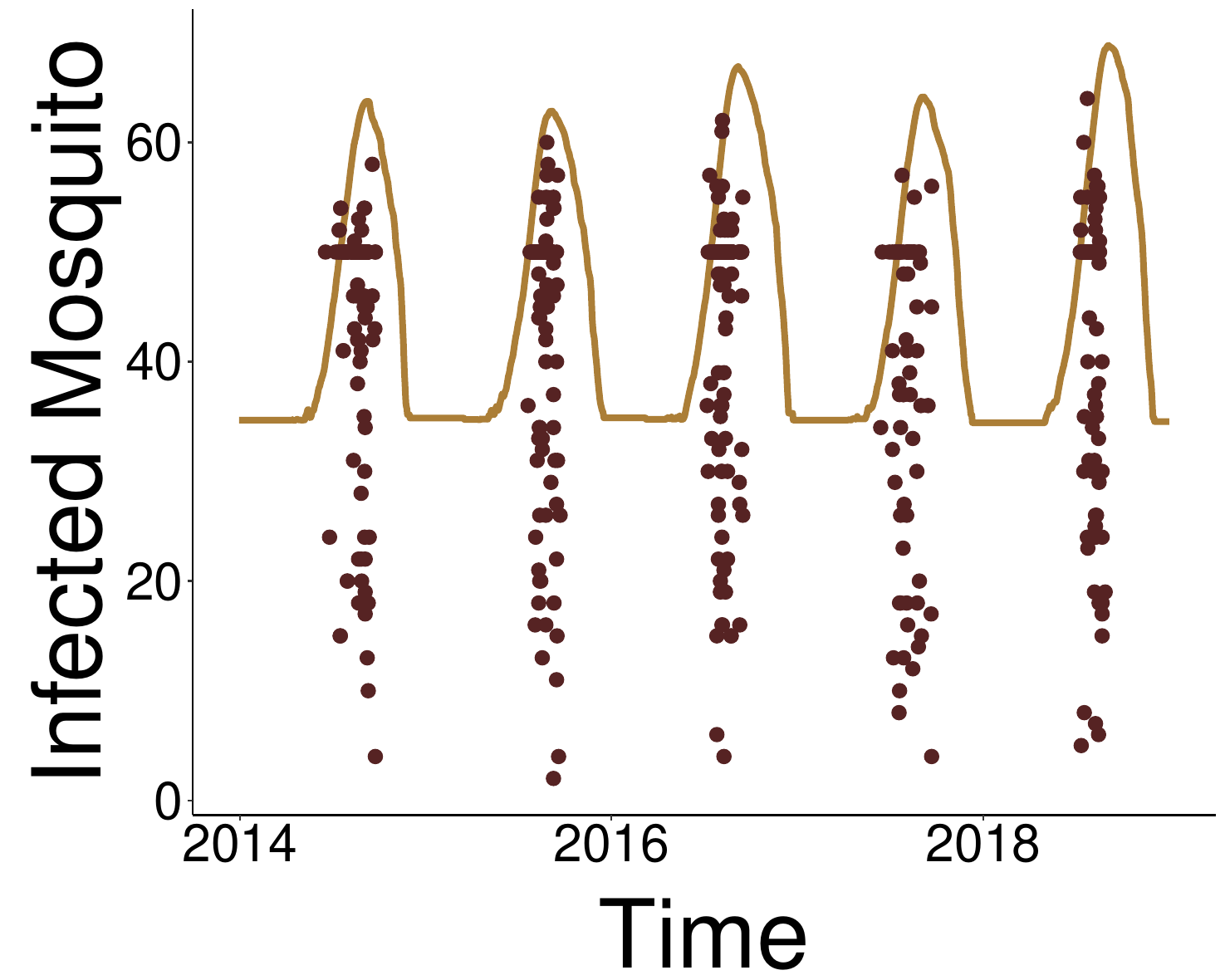}
\caption{Simulated infected mosquito abundance  as described in  \eqref{Eq1mos} and the trap data collected from the cook county, Illinois. 
Entomological observations made by trapping are represented by dark brown circles.
These circles represent the number WNV positive adult mosquitoes.
The simulated abundance of infected mosquito modelled as detailed in \eqref{Eq1mos} is depicted by the yellow line in each trap location. 
Period is between 2014–2018.
}
    \label{fig:ModelValid1}
\end{figure}

\begin{figure}[H]
\centering
\includegraphics[width=8cm]{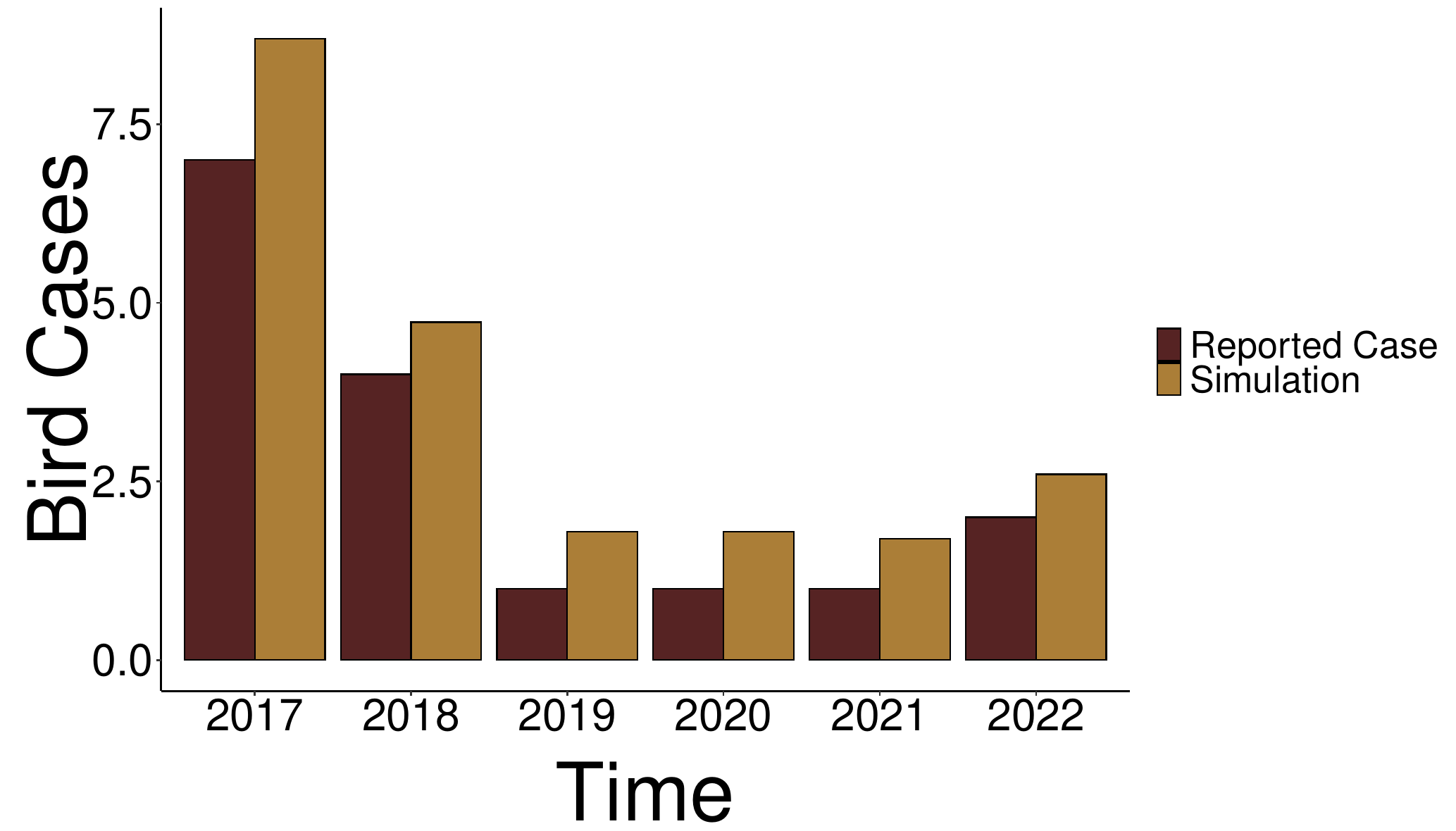}
\caption{Simulated number of infected bird described in \eqref{Eq3bir} and the reported data \cite{IDPH}.
We  scale our model simulations to calibrate to match the overall sums in reported data.
Period is between 2017–2022.
}
    \label{fig:ModelValid2}
\end{figure}

\begin{figure}[H]
\centering
\includegraphics[width=8cm]{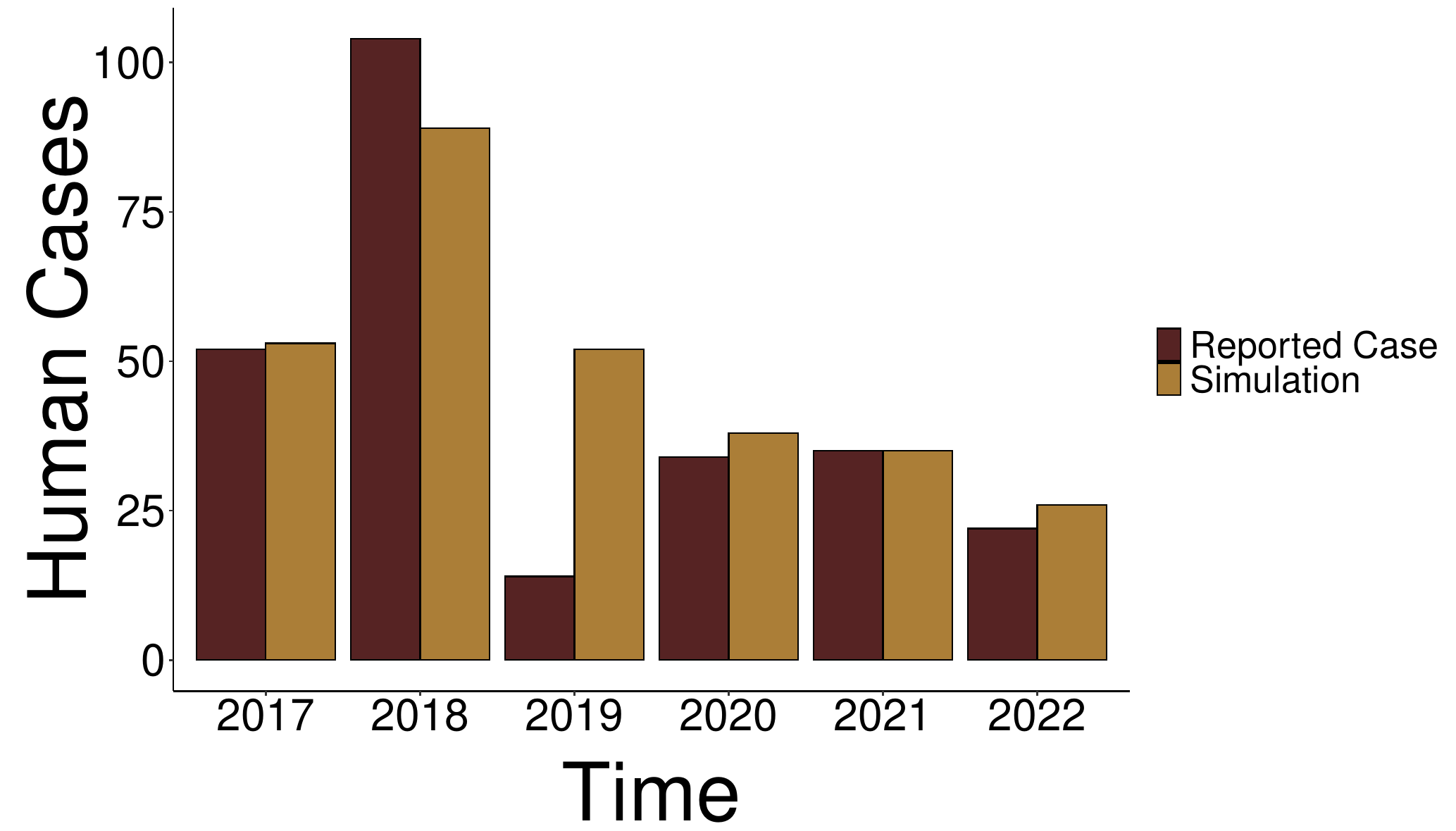}
\caption{Simulated WNV infected human  described in  \eqref{Eq34hum} and the reported data \cite{IDPH}.
We also calibrate our model output scale to adjust ensuring that the simulation outputs align with the observed aggregate values.
Period is between 2017–2022.}
    \label{fig:ModelValid3}
\end{figure}

Figure \ref{fig:ModelValid1} presents a comparison between the observed number of WNV positive mosquitoes in the trap data and simulated cases of WNV positive mosquitoes. 
We also  compare the reported and simulated bird and human cases of WNV in Figure \ref{fig:ModelValid2} and 
Figure \ref{fig:ModelValid3}.
To validate the accuracy of our model simulations (\eqref{Eq3bir}, \eqref{Eq34hum}), we  aggregate our daily model outputs into yearly values to align with the available reported data. 
In addition, we scale the simulated model outputs to match with the magnitude of the reported data by calibrating the sums of the reported and simulations.
We  scale the magnitude because only a proportion of the actual WNV cases is reported that are available to us.
The pursuit of this verification process is to assess how well the predictions from the mathematical model are aligned with the reported cases of WNV across different interacting populations.\par
We find a higher correlation value for humans ($R = 0.82$) compared to the reported cases of WNV infected birds ($R = 0.73$).
Through this we are able to assess the strength of the relationship amongst the variables being compared. 
We can surmise that the higher correlation found in case of human cases due to better reliability of surveillance data for humans compared to birds. 
In other words, the reported data of human cases of WNV perhaps more accurate and consistent, thus leading to a higher value of $R$.
However, we acknowledge that these values of $R$, perhaps, do not necessarily suggest independent model verification. 
Alternately, we would like to emphasise that these values of $R$ actually demonstrate the strength of our mathematical model (\eqref{Eq3bir}, \eqref{Eq34hum}) to capture the trend in the reported cases of WNV and replicate the WNV transmission dynamics.

\section{Discussion and Conclusion}
%\textbf{\textcolor{blue}{Our model and the findings}}\par
We develope a compartment model that includes feeding preference of the vector species to understand the local spread of WNV transmission. 
This model incorporates various compartments representing different interacting population, such as susceptible individuals, infected individuals,  exposed individuals and recovered individuals. 
By considering the interactions amongst these compartments with temperature-driven model parameters, the model can simulate the spread of the infection in a local habitat. 
Additionally, the model allows for the inclusion of various influencing factors that impact the transmission dynamics of WNV, such as mosquito abundance, different host population sizes, various environmental factors, and intervention measures. 
In our modelling study, we also  investigate the impact of heterogeneity in contact rates amongst vectors (mosquitoes) and different hosts on the dynamics of WNV transmission in pertinent to the feeding index of the vector species in a climate-driven ODE system.
To answer this question, we incorporate the feeding index that was introduced in \cite{doi:10.1098/rspb.2011.1282} and additionally we also include the time varying introduction of infected agents (birds) and these are accounted for the heterogeneity in contact rates, while keeping the model parameters weather driven.
By simulating the WNV transmission model under various scenarios, we are able to assess the potential influence.\par 

Our current modelling effort improves upon previous modelling attempts by incorporating both including the weather-dependent parametrisation and validation of all the interacting species. 
We develope a generic ODE based model that is flexible enough to incorporate multiple species, further compartments to accommodate different health status of hosts-vector and various weather-dependent parameters.
This approach offers the benefits of a robust, versatile model that is site independent and it can be extended to investigate transmission dynamics within and between multiple localities.\par
%\textbf{\textcolor{blue}{Similarities with other work}}\par
Consistent with previous research efforts \cite{doi:10.1098/rspb.2011.1282,  doi:10.3920/978-90-8686-932-9_12}, simulations clearly indicate that there is a greater influence of $\alpha_F$ on the transmission dynamics of WNV (Figures \ref{fig:alphaFetavsIB}). 
For example, the authors in \cite{doi:10.1098/rspb.2011.1282},  demonstrate that beyond a threshold value of $\alpha_F$, the WNV transmission dynamics undergoes a radical shift and we explore it in a great detail.
In Figure \ref{fig:alphaVsR0} we demonstrate the phase transition from no outbreak to a WNV outbreak and we are able to establish the functional relationship between $R_0$ and $\alpha_F$ that governs the WNV outbreak situation. 
Similar to the findings in \cite{doi:10.1098/rspb.2011.1282}, $\alpha_F$ is found to be one of the most sensitive parameters and its influence on the WNV dynamics is very important to note. 

%\textbf{\textcolor{blue}{Novelty of our work}}\par

%\textbf{\textcolor{blue}{Conclusion}}\par
\par
Through our modelling effort, we aim to provide insights into the factors driving the spread of WNV and inform strategies for controlling and mitigating its impact on public health.

\section{Acknowledgement}
This publication was supported by Cooperative Agreement Number U01CK000651 from the Centers for Disease Control and Prevention. 
Its contents are solely the responsibility of the authors and do not necessarily represent the official views of the Centers for Disease Control and Prevention.

\bibliography{Mybib}

\begin{thebibliography}{10}
\expandafter\ifx\csname url\endcsname\relax
  \def\url#1{\texttt{#1}}\fi
\expandafter\ifx\csname urlprefix\endcsname\relax\def\urlprefix{URL }\fi
\expandafter\ifx\csname href\endcsname\relax
  \def\href#1#2{#2} \def\path#1{#1}\fi

\bibitem{doi:10.1126/science.1201010}
A.~M. Kilpatrick, Globalization, land use, and the invasion of west nile virus,
  Science 334~(6054) (2011) 323--327.

\bibitem{doi:10.1128/jvi.01963-10}
F.~J. May, C.~T. Davis, R.~B. Tesh, A.~D.~T. Barrett, Phylogeography of west
  nile virus: from the cradle of evolution in africa to eurasia, australia, and
  the americas, Journal of Virology 85~(6) (2011) 2964--2974.

\bibitem{doi:10.1128/cmr.00045-12}
T.~M. Colpitts, M.~J. Conway, R.~R. Montgomery, E.~Fikrig, West nile virus:
  Biology, transmission, and human infection, Clinical Microbiology Reviews
  25~(4) (2012) 635--648.

\bibitem{10.1093/auk/124.4.1121}
A.~M. Kilpatrick, S.~L. LaDeau, P.~P. Marra, {Ecology of West Nile Virus
  Transmission and its Impact on Birds in the Western Hemisphere}, The Auk
  124~(4) (2007) 1121--1136.

\bibitem{KRAMER2007171}
L.~D. Kramer, J.~Li, P.-Y. Shi, West nile virus, The Lancet Neurology 6~(2)
  (2007) 171--181.

\bibitem{10.1093/jme/tjz151}
L.~D. Kramer, A.~T. Ciota, A.~M. Kilpatrick, {Introduction, Spread, and
  Establishment of West Nile Virus in the Americas}, Journal of Medical
  Entomology 56~(6) (2019) 1448--1455.

\bibitem{refId0}
{Murray, Kristy O.}, {Mertens, Eva}, {Despr\`es, Philippe}, West nile virus and
  its emergence in the united states of america, Vet. Res. 41~(6) (2010) 67.

\bibitem{Allan}
B.~F. Allan, R.~B. Langerhans, W.~A. Ryberg, W.~J. Landesman, N.~W. Griffin,
  R.~S. Katz, B.~J. Oberle, M.~R. Schutzenhofer, K.~N. Smyth,
  A.~de~St.~Maurice, L.~Clark, K.~R. Crooks, D.~E. Hernandez, R.~G. McLean,
  R.~S. Ostfeld, J.~M. Chase, Ecological correlates of risk and incidence of
  west nile virus in the united states, Oecologia 158~(4) (2009) 699--708.

\bibitem{https://doi.org/10.1111/gcb.15842}
A.~C. Keyel, A.~Raghavendra, A.~T. Ciota, O.~Elison~Timm, West nile virus is
  predicted to be more geographically widespread in new york state and
  connecticut under future climate change, Global Change Biology 27~(21) (2021)
  5430--5445.

\bibitem{doi:10.1289/ehp.0800487}
J.~E. Soverow, G.~A. Wellenius, D.~N. Fisman, M.~A. Mittleman, Infectious
  disease in a warming world: How weather influenced west nile virus in the
  united states (2001--2005), Environmental Health Perspectives 117~(7) (2009)
  1049--1052.

\bibitem{https://doi.org/10.1029/2022GH000708}
H.~M. Hort, M.~Ibaraki, F.~W. Schwartz, Temporal and spatial synchronicity in
  west nile virus cases along the central flyway, usa, GeoHealth 7~(5) (2023)
  e2022GH000708.

\bibitem{doi:10.1098/rspb.2006.3575}
A.~Marm~Kilpatrick, P.~Daszak, M.~J. Jones, P.~P. Marra, L.~D. Kramer, Host
  heterogeneity dominates west nile virus transmission, Proceedings of the
  Royal Society B: Biological Sciences 273~(1599) (2006) 2327--2333.

\bibitem{doi:10.1098/rspb.2011.1282}
J.~E. Simpson, P.~J. Hurtado, J.~Medlock, G.~Molaei, T.~G. Andreadis, A.~P.
  Galvani, M.~A. Diuk-Wasser, Vector host-feeding preferences drive
  transmission of multi-host pathogens: West nile virus as a model system,
  Proceedings of the Royal Society B: Biological Sciences 279~(1730) (2012)
  925--933.

\bibitem{Levine}
R.~S. Levine, D.~L. Hedeen, M.~W. Hedeen, G.~L. Hamer, D.~G. Mead, U.~D.
  Kitron, Avian species diversity and transmission of west nile virus in
  atlanta, georgia, Parasites \& Vectors 10~(1) (2017) 62.

\bibitem{10.1371/journal.pone.0039549}
J.~Mu{\~n}oz, S.~Ruiz, R.~Soriguer, M.~Alcaide, D.~S. Viana, D.~Roiz,
  A.~V{\'a}zquez, J.~Figuerola, Feeding patterns of potential west nile virus
  vectors in south-west spain, PLOS ONE 7~(6) (2012) 1--9.

\bibitem{Sarah}
S.~S. Wheeler, C.~C. Taff, W.~K. Reisen, A.~K. Townsend, Mosquito blood-feeding
  patterns and nesting behavior of american crows, an amplifying host of west
  nile virus, Parasites \& Vectors 14~(1) (2021) 331.

\bibitem{HostSelection}
G.~L. Hamer, U.~D. Kitron, T.~L. Goldberg, J.~D. Brawn, S.~R. Loss, M.~O. Ruiz,
  D.~B. Hayes, E.~D. Walker, Host selection by culex pipiens mosquitoes and
  west nile virus amplification, The American Journal of Tropical Medicine and
  Hygiene Am J Trop Med Hyg 80~(2) (2009) 268 -- 278.

\bibitem{Komar}
N.~Komar, J.~M. Colborn, K.~Horiuchi, M.~Delorey, B.~Biggerstaff, D.~Damian,
  K.~Smith, J.~Townsend, Reduced west nile virus transmission around communal
  roosts of great-tailed grackle (quiscalus mexicanus), EcoHealth 12~(1) (2015)
  144--151.

\bibitem{doi:10.3920/978-90-8686-932-9_12}
N.~Stanczyk, C.~D. Moraes, M.~Mescher, Chapter 12: Effects of pathogens on
  mosquito host-seeking and feeding behaviour, Ch.~12, pp. 327--348.

\bibitem{ROHR2011270}
J.~R. Rohr, A.~P. Dobson, P.~T. Johnson, A.~M. Kilpatrick, S.~H. Paull, T.~R.
  Raffel, D.~Ruiz-Moreno, M.~B. Thomas, Frontiers in climate change--disease
  research, Trends in Ecology \& Evolution 26~(6) (2011) 270--277.

\bibitem{POH2019260}
K.~C. Poh, L.~F. Chaves, M.~Reyna-Nava, C.~M. Roberts, C.~Fredregill, R.~Bueno,
  M.~Debboun, G.~L. Hamer, The influence of weather and weather variability on
  mosquito abundance and infection with west nile virus in harris county,
  texas, usa, Science of The Total Environment 675 (2019) 260--272.

\bibitem{10.7554/eLife.58511}
M.~S. Shocket, A.~B. Verwillow, M.~G. Numazu, H.~Slamani, J.~M. Cohen,
  F.~El~Moustaid, J.~Rohr, L.~R. Johnson, E.~A. Mordecai, Transmission of west
  nile and five other temperate mosquito-borne viruses peaks at temperatures
  between 23$\,^{\circ}$c and 26$\,^{\circ}$c, eLife 9 (2020) e58511.

\bibitem{Ruiz1}
M.~O. Ruiz, L.~F. Chaves, G.~L. Hamer, T.~Sun, W.~M. Brown, E.~D. Walker,
  L.~Haramis, T.~L. Goldberg, U.~D. Kitron, Local impact of temperature and
  precipitation on west nile virus infection in culex species mosquitoes in
  northeast illinois, usa, Parasites \& Vectors 3~(1) (2010) 19.

\bibitem{doi:10.1098/rsos.170017}
Y.~Wang, W.~Pons, J.~Fang, H.~Zhu, The impact of weather and storm water
  management ponds on the transmission of west nile virus, Royal Society Open
  Science 4~(8) (2017) 170017.

\bibitem{10.1371/journal.pone.0161510}
N.~I. Stilianakis, V.~Syrris, T.~Petroliagkis, P.~P{\"a}rt, S.~Gewehr,
  S.~Kalaitzopoulou, S.~Mourelatos, A.~Baka, D.~Pervanidou, J.~Vontas,
  C.~Hadjichristodoulou, Identification of climatic factors affecting the
  epidemiology of human west nile virus infections in northern greece, PLOS ONE
  11~(9) (2016) 1--17.

\bibitem{10.1093/jmedent/43.2.309}
W.~K. Reisen, Y.~Fang, V.~M. Martinez, {Effects of Temperature on the
  Transmission of West Nile Virus by Culex tarsalis (Diptera: Culicidae) },
  Journal of Medical Entomology 43~(2) (2014) 309--317.

\bibitem{FAY2022147}
R.~L. Fay, A.~C. Keyel, A.~T. Ciota, Chapter three - west nile virus and
  climate change, in: M.~J. Roossinck (Ed.), Viruses and Climate Change, Vol.
  114 of Advances in Virus Research, Academic Press, 2022, pp. 147--193.

\bibitem{GIESEN2023100478}
C.~Giesen, Z.~Herrador, B.~Fernandez-Martinez, J.~Figuerola, L.~Gangoso,
  A.~Vazquez, D.~G{\'o}mez-Barroso, A systematic review of environmental
  factors related to wnv circulation in european and mediterranean countries,
  One Health 16 (2023) 100478.

\bibitem{10.1371/journal.pone.0001146}
K.~P. Paaijmans, M.~O. Wandago, A.~K. Githeko, W.~Takken, Unexpected high
  losses of anopheles gambiae larvae due to rainfall, PLOS ONE 2~(11) (2007)
  1--7.

\bibitem{10.1371/journal.pntd.0006935}
C.~M. Benedum, O.~M.~E. Seidahmed, E.~A.~B. Eltahir, N.~Markuzon, Statistical
  modeling of the effect of rainfall flushing on dengue transmission in
  singapore, PLOS Neglected Tropical Diseases 12~(12) (2018) 1--18.

\bibitem{https://doi.org/10.1111/ele.14228}
J.~J. Brown, M.~Pascual, M.~C. Wimberly, L.~R. Johnson, C.~C. Murdock, Humidity
  -- the overlooked variable in the thermal biology of mosquito-borne disease,
  Ecology Letters 26~(7) (2023) 1029--1049.

\bibitem{10.1371/journal.pcbi.1006047}
N.~B. DeFelice, Z.~D. Schneider, E.~Little, C.~Barker, K.~A. Caillouet, S.~R.
  Campbell, D.~Damian, P.~Irwin, H.~M.~P. Jones, J.~Townsend, J.~Shaman, Use of
  temperature to improve west nile virus forecasts, PLOS Computational Biology
  14~(3) (2018) 1--25.

\bibitem{BHOWMICK2020110117}
S.~Bhowmick, J.~Gethmann, F.~J. Conraths, I.~M. Sokolov, H.~H. Lentz, Locally
  temperature - driven mathematical model of west nile virus spread in germany,
  Journal of Theoretical Biology 488 (2020) 110117.

\bibitem{LAPERRIERE201199}
V.~Laperriere, K.~Brugger, F.~Rubel, Simulation of the seasonal cycles of bird,
  equine and human west nile virus cases, Preventive Veterinary Medicine 98~(2)
  (2011) 99--110.

\bibitem{10.1371/journal.pone.0246046}
A.~B.~B. Wilke, C.~Vasquez, A.~Carvajal, M.~Ramirez, G.~Cardenas, W.~D. Petrie,
  J.~C. Beier, Effectiveness of adulticide and larvicide in controlling high
  densities of aedes aegypti in urban environments, PLOS ONE 16~(1) (2021)
  1--15.

\bibitem{10.1093/jme/tjad088}
K.~Lopez, P.~Irwin, G.~M. Bron, S.~Paskewitz, L.~Bartholomay, {Ultra-low volume
  (ULV) adulticide treatment impacts age structure of Culex species (Diptera:
  Culicidae) in a West Nile virus hotspot}, Journal of Medical Entomology
  60~(5) (2023) 1108--1116.

\bibitem{Vega}
M.~Santos-Vega, P.~P. Martinez, K.~G. Vaishnav, V.~Kohli, V.~Desai, M.~J.
  Bouma, M.~Pascual, The neglected role of relative humidity in the interannual
  variability of urban malaria in indian cities, Nature Communications 13~(1)
  (2022) 533.

\bibitem{Klump}
M.~Klumpp, D.~Loske, S.~Bicciato, Covid-19 health policy evaluation:
  integrating health and economic perspectives with a data envelopment analysis
  approach, The European Journal of Health Economics 23~(8) (2022) 1263--1285.

\bibitem{Bergsman}
L.~D. Bergsman, J.~M. Hyman, C.~A. Manore, A mathematical model for the spread
  of west nile virus in migratory and resident birds, Mathematical Biosciences
  and Engineering 13~(2) (2016) 401--424.

\bibitem{BHOWMICK2023110213}
S.~Bhowmick, J.~Gethmann, F.~J. Conraths, I.~M. Sokolov, H.~H. Lentz,
  Seir-metapopulation model of potential spread of west nile virus, Ecological
  Modelling 476 (2023) 110213.

\bibitem{BHOWMICK2023104827}
S.~Bhowmick, I.~M. Sokolov, H.~H. Lentz, Decoding the double trouble: A
  mathematical modelling of co-infection dynamics of sars-cov-2 and
  influenza-like illness, Biosystems 224 (2023) 104827.

\bibitem{BOWMAN20051107}
C.~Bowman, A.~Gumel, P.~{van den Driessche}, J.~Wu, H.~Zhu, A mathematical
  model for assessing control strategies against west nile virus, Bulletin of
  Mathematical Biology 67~(5) (2005) 1107--1133.

\bibitem{10.1371/journal.pone.0227160}
S.~Karki, W.~M. Brown, J.~Uelmen, M.~O. Ruiz, R.~L. Smith, The drivers of west
  nile virus human illness in the chicago, illinois, usa area: Fine scale
  dynamic effects of weather, mosquito infection, social, and biological
  conditions, PLOS ONE 15~(5) (2020) 1--19.

\bibitem{EffectsofScaleonModelingWestNileVirusDiseaseRisk}
J.~A. Uelmen, P.~Irwin, D.~Bartlett, W.~Brown, S.~Karki, M.~O. Ruiz,
  J.~Fraterrigo, B.~Li, R.~L. Smith, Effects of scale on modeling west nile
  virus disease risk, The American Journal of Tropical Medicine and Hygiene
  104~(1) (2021) 151 -- 165.

\bibitem{doi:10.1289/EHP10287}
M.~C. Wimberly, J.~K. Davis, M.~B. Hildreth, J.~L. Clayton, Integrated
  forecasts based on public health surveillance and meteorological data predict
  west nile virus in a high-risk region of north america, Environmental Health
  Perspectives 130~(8) (2022) 087006.

\bibitem{10.1371/journal.pntd.0010252}
E.~Fesce, G.~Marini, R.~Ros{\`a}, D.~Lelli, M.~P. Cerioli, M.~Chiari,
  M.~Farioli, N.~Ferrari, Understanding west nile virus transmission:
  Mathematical modelling to quantify the most critical parameters to predict
  infection dynamics, PLOS Neglected Tropical Diseases 17~(5) (2023) 1--21.

\bibitem{https://doi.org/10.1111/nrm.12165}
E.~Schaefer, K.~A. Caillou{\"e}t, S.~L. Robertson, Methods for prophylactic
  management of west nile virus using a stage-structured avian host-vector
  model with vaccination, larvicide, and adulticide, Natural Resource Modeling
  31~(4) (2018) e12165.

\bibitem{10.1371/journal.pone.0108452}
K.~A. Pawelek, P.~Niehaus, C.~Salmeron, E.~J. Hager, G.~J. Hunt, Modeling
  dynamics of culex pipiens complex populations and assessing abatement
  strategies for west nile virus, PLOS ONE 9~(9) (2014) 1--15.

\bibitem{10.3389/fevo.2022.993844}
L.~Chen, S.~Chen, P.~Kong, L.~Zhou, Host competence, interspecific competition
  and vector preference interact to determine the vector-borne infection
  ecology, Frontiers in Ecology and Evolution 10 (2022).

\bibitem{MALIK201860}
T.~Malik, A discrete time west nile virus transmission model with optimal bird-
  and vector-specific controls, Mathematical Biosciences 305 (2018) 60--70.

\bibitem{Demers}
J.~Demers, S.~L. Robertson, S.~Bewick, W.~F. Fagan, Implicit versus explicit
  vector management strategies in models for vector-borne disease epidemiology,
  Journal of Mathematical Biology 84~(6) (2022) 48.

\bibitem{mbs:/content/journal/jgv/10.1099/vir.0.033829-0}
E.~Sotelo, J.~Fern{\'a}ndez-Pinero, F.~Llorente, A.~V{\'a}zquez, A.~Moreno,
  M.~Ag{\"u}ero, P.~Cordioli, A.~Tenorio, M.~{\'A}. Jim{\'e}nez-Clavero,
  Phylogenetic relationships of western mediterranean west nile virus strains
  (1996--2010) using full-length genome sequences: single or multiple
  introductions?, Journal of General Virology 92~(11) (2011) 2512--2522.

\bibitem{Mann}
B.~Mann, A.~McMullen, D.~Swetnam, V.~Salvato, M.~Reyna, H.~Guzman, R.~Bueno,
  J.~Dennett, R.~Tesh, A.~D.~T. Barrett, Continued evolution of west nile
  virus, houston, texas, usa, 2002--2012, Emerging Infectious Disease journal
  19~(9) (2013) 1418.

\bibitem{doi:10.1098/rstb.2010.0054}
G.~Amore, L.~Bertolotti, G.~L. Hamer, U.~D. Kitron, E.~D. Walker, M.~O. Ruiz,
  J.~D. Brawn, T.~L. Goldberg, Multi-year evolutionary dynamics of west nile
  virus in suburban chicago, usa, 2005--2007, Philosophical Transactions of the
  Royal Society B: Biological Sciences 365~(1548) (2010) 1871--1878.

\bibitem{IDPH}
\href{https://dph.illinois.gov/topics-services/diseases-and-conditions/west-nile-virus/surveillance.html}{Illinois
  department of public health}, accessed: 2023-06-30.
\newline\urlprefix\url{https://dph.illinois.gov/topics-services/diseases-and-conditions/west-nile-virus/surveillance.html}

\bibitem{PRISMClimateGroup}
\href{http://prism.oregonstate.edu}{Prism climate group, oregon state
  university}, accessed: 2023-04-30.
\newline\urlprefix\url{http://prism.oregonstate.edu}

\bibitem{RUBEL2008166}
F.~Rubel, K.~Brugger, M.~Hantel, S.~Chvala-Mannsberger, T.~Bakonyi,
  H.~Weissenb{\"o}ck, N.~Nowotny, Explaining usutu virus dynamics in austria:
  Model development and calibration, Preventive Veterinary Medicine 85~(3)
  (2008) 166--186.

\bibitem{Grub}
N.~D. Grubaugh, J.~T. Ladner, M.~U.~G. Kraemer, G.~Dudas, A.~L. Tan,
  K.~Gangavarapu, M.~R. Wiley, S.~White, J.~Th{\'e}z{\'e}, D.~M. Magnani,
  K.~Prieto, D.~Reyes, A.~M. Bingham, L.~M. Paul, R.~Robles-Sikisaka,
  G.~Oliveira, D.~Pronty, C.~M. Barcellona, H.~C. Metsky, M.~L. Baniecki, K.~G.
  Barnes, B.~Chak, C.~A. Freije, A.~Gladden-Young, A.~Gnirke, C.~Luo,
  B.~MacInnis, C.~B. Matranga, D.~J. Park, J.~Qu, S.~F. Schaffner,
  C.~Tomkins-Tinch, K.~L. West, S.~M. Winnicki, S.~Wohl, N.~L. Yozwiak,
  J.~Quick, J.~R. Fauver, K.~Khan, S.~E. Brent, R.~C. Reiner, P.~N.
  Lichtenberger, M.~J. Ricciardi, V.~K. Bailey, D.~I. Watkins, M.~R. Cone,
  E.~W. Kopp, K.~N. Hogan, A.~C. Cannons, R.~Jean, A.~J. Monaghan, R.~F. Garry,
  N.~J. Loman, N.~R. Faria, M.~C. Porcelli, C.~Vasquez, E.~R. Nagle, D.~A.~T.
  Cummings, D.~Stanek, A.~Rambaut, M.~Sanchez-Lockhart, P.~C. Sabeti, L.~D.
  Gillis, S.~F. Michael, T.~Bedford, O.~G. Pybus, S.~Isern, G.~Palacios, K.~G.
  Andersen, Genomic epidemiology reveals multiple introductions of zika virus
  into the united states, Nature 546~(7658) (2017) 401--405.

\bibitem{Pet}
M.~E. Petrone, R.~Earnest, J.~Louren{\c c}o, M.~U.~G. Kraemer,
  R.~Paulino-Ramirez, N.~D. Grubaugh, L.~Tapia, Asynchronicity of endemic and
  emerging mosquito-borne disease outbreaks in the dominican republic, Nature
  Communications 12~(1) (2021) 151.

\bibitem{doi:10.1098/rsif.2009.0386}
O.~Diekmann, J.~A.~P. Heesterbeek, M.~G. Roberts, The construction of
  next-generation matrices for compartmental epidemic models, Journal of The
  Royal Society Interface 7~(47) (2010) 873--885.

\bibitem{VANDENDRIESSCHE200229}
P.~{van den Driessche}, J.~Watmough, Reproduction numbers and sub-threshold
  endemic equilibria for compartmental models of disease transmission,
  Mathematical Biosciences 180~(1) (2002) 29--48.

\bibitem{sensivity}
B.~Iooss, S.~D. Veiga, A.~Janon, G.~Pujol, B.~Broto, K.~Boumhaout, L.~Clouvel,
  T.~Delage, R.~E. Amri, J.~Fruth, L.~Gilquin, J.~Guillaume, M.~Herin, M.~I.
  Idrissi, L.~{Le Gratiet}, P.~Lemaitre, A.~Marrel, A.~Meynaoui, B.~L. Nelson,
  F.~Monari, R.~Oomen, O.~Rakovec, B.~Ramos, O.~Roustant, G.~Sarazin, E.~Song,
  J.~Staum, R.~Sueur, T.~Touati, V.~Verges, F.~Weber,
  \href{https://CRAN.R-project.org/package=sensitivity}{sensitivity: Global
  Sensitivity Analysis of Model Outputs}, r package version 1.29.0 (2023).
\newline\urlprefix\url{https://CRAN.R-project.org/package=sensitivity}

\bibitem{lhs}
R.~Carnell, \href{https://github.com/bertcarnell/lhs}{lhs: Latin Hypercube
  Samples}, r package version 1.1.5 (2022).
\newline\urlprefix\url{https://github.com/bertcarnell/lhs}

\bibitem{R}
{R Core Team}, \href{https://www.R-project.org/}{R: A Language and Environment
  for Statistical Computing}, R Foundation for Statistical Computing, Vienna,
  Austria (2021).
\newline\urlprefix\url{https://www.R-project.org/}

\end{thebibliography}
\end{document}